\documentclass[
    aps,
    physrev,
    twocolumn,
    groupedaddress,
    amssymb,
    amsmath,
    10pt,
]{revtex4-2}
\usepackage{graphicx}
\usepackage{bm}
\usepackage{xcolor} 
\usepackage{mathtools} 
\usepackage[normalem]{ulem}

\begin{document}


\title{Critical dephasing rates for the observation of collective behavior in a pair of coupled quantum emitters}


\author{Sébastien Quistrebert}
\author{Elisabeth Gliott}
\author{Jean-Sébastien Lauret}
\author{Nikos Fayard}
\email[nikos.fayard@ens-paris-saclay.fr]{}
\affiliation{ Université Paris-Saclay, CNRS, Ecole Normale Supérieure Paris-Saclay, CentraleSupélec, LuMIn, Orsay 91190, France}


\date{\today}
\begin{abstract}
Efficient atom-photon interfaces require the controlled assembly of quantum emitters, where collective effects such as superradiance and subradiance can emerge. Recent experiments with subwavelength arrays of quantum dots have reported superradiance at room temperature~\cite{huang2022room,biliroglu2022room}, revealing a delicate competition between collective enhancement of coherent emission and pure dephasing $\gamma^*$, which destroys it. Motivated by these results, we theoretically study $N=2$ coupled quantum emitters and identify threshold values of $\gamma^*$, for four experimentally accessible observables, beyond which collective effects vanish. The thresholds depend sensitively on the chosen observable, highlighting the subtlety of detecting collective behavior. Our work provides a quantitative framework to guide experiments and optimize conditions for observing collective quantum phenomena.
\end{abstract}


\maketitle


\section{Introduction}
The combination of atoms’ intrinsic ability to store and process quantum information with the long-distance, low-loss propagation of photons establishes atom–photon interfaces as leading candidates for realizing scalable quantum networks~\cite{hammerer2010quantum,raimond2001manipulating,saffman2010quantum}.
To fully exploit the potential of such platforms~\cite{kimble2008quantum}, it is essential to strengthen light-matter interactions to maximize atom-photon coupling~\cite{chang2018colloquium}, while simultaneously minimizing the various sources of decoherence within the system~\cite{haroche1998entanglement,lidar1998decoherence}.

There are several ways to enhance light-matter interactions.
One approach involves engineering the photonic density of states by interfacing emitters with nanostructures such as optical cavities~\cite{thompson2013coupling,reiserer2015cavity,plankensteiner2017cavity,shlesinger2021time,lei2023many}, nanofibers~\cite{nayak2007optical,solano2017super,corzo2019waveguide,pennetta2022collective,jimenez2025controlling}, or photonic crystal waveguides~\cite{goban2014atom,goban2015superradiance,lodahl2015interfacing,yu2014nanowire}. In such systems, the Purcell effect can lead to quasi-one-dimensional spontaneous emission~\cite{turchette1995measurement,chang2014quantum,roy2017colloquium,sheremet2023waveguide}. 
However, these platforms remain challenging to develop, mainly due to the difficulties in efficiently coupling a large number of atoms to nanostructures~\cite{sheremet2023waveguide}.
A second approach leverages collective effects. Recent studies have shown that strong light-matter interactions can emerge in large atomic arrays assembled in free space~\cite{bettles2016enhanced,facchinetti2016storing,shahmoon2017cooperative,rui2020subradiant}. When the interatomic spacing is reduced below the optical wavelength, light-mediated interactions between atoms become significant enough to give rise to collective phenomena that modify the optical response of the system. This can manifest as superradiance~\cite{dicke1954coherence,gross1982superradiance,scully2006directed,araujo2016superradiance,he2020atomic,he2020geometric,rastogi2022superradiance}, characterized by an enhanced emission rate of collective modes, or subradiance~\cite{scully2015single,plankensteiner2015selective,guerin2016subradiance,cipris2021subradiance,ferioli2021storage}, where the emission rate is suppressed.

\begin{figure*}
\includegraphics[width=\textwidth]{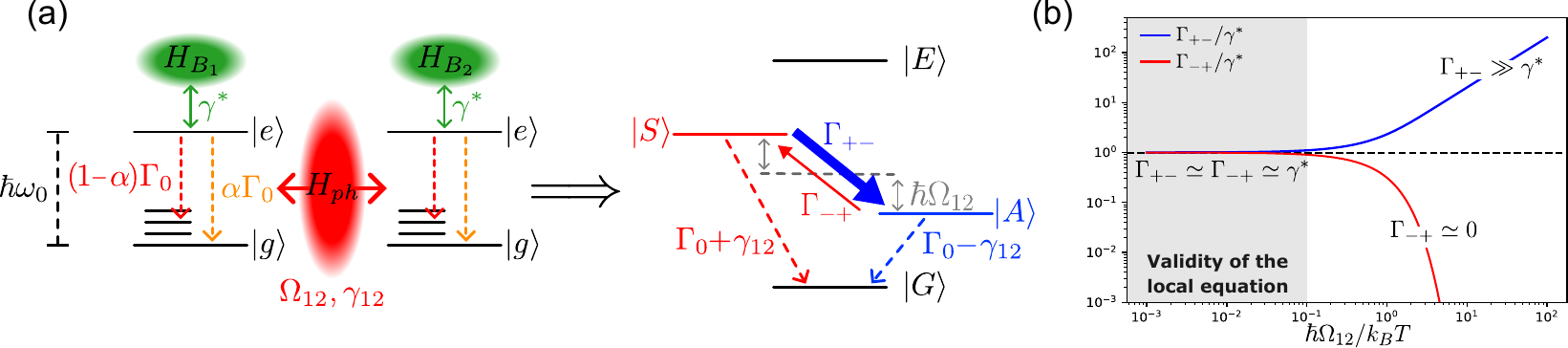}
\caption{\label{fig:intro} (a) Representation of the system under study: a pair of two-level systems coupled through a common photonic bath and each interacting with an independent dephasing bath. The vibrational levels shown above the ground state illustrate the phonon-assisted emission processes typical of solid-state emitters, with a fraction $1-\alpha$ of the spontaneous emission being red-shifted. 
The corresponding Hamiltonian eigenstates of the coupled system are shown on the right. For two two-level emitters, the Hilbert space consists of four states ($2^2=4$): the ground state $|G\rangle$, the symmetric and antisymmetric collective states $|S\rangle$ and $|A\rangle$, and the doubly excited state $|E\rangle$. The diagram also indicates the radiative decay rates and the dephasing-induced transition rates $\Gamma_{+-}$ and $\Gamma_{-+}$ between the collective states $|S\rangle$ and $|A\rangle$ introduced in Eqs.~\eqref{eq:transition_rate_SA}--\eqref{eq:transition_rate_AS}. (b) Temperature dependence of the dephasing-induced transition rates, normalized to the local dephasing rate $\gamma^*$, as a function of the dimensionless ratio $\hbar|\Omega_{12}|/k_BT$. In the high-temperature regime $\hbar|\Omega_{12}|/k_BT\ll1$, both rates approach $\gamma^*$, corresponding to the local master equation. As the ratio increases, the two rates progressively separate, and the global description becomes increasingly relevant.}\end{figure*}

From an experimental point of view, the ability to create large arrays of $N$ neutral atoms with $\mu m$  interatomic distances has already been demonstrated in order to perform quantum simulations ~\cite{bloch2012quantum,nogrette2014single,endres2016atom,barredo2016atom}. 
The reduction of the atomic distance below the wavelength (which is crucial to access subradiant states) is extremely challenging for neutral atoms since the trapping is done optically~\cite{rui2020subradiant,srakaew2023subwavelength}.
In contrast, nano-objects such as molecules, semiconductor nanocrystals or quantum dots (QD)  can be placed close together at distances  one or two orders of magnitudes smaller than the wavelength of visible light. Moreover, being composed of dozens to thousands of atoms, they often exhibit large transition dipole moments, which opens up the possibility for strong light-matter interactions. 

\begin{figure*}
\includegraphics[width=\textwidth]{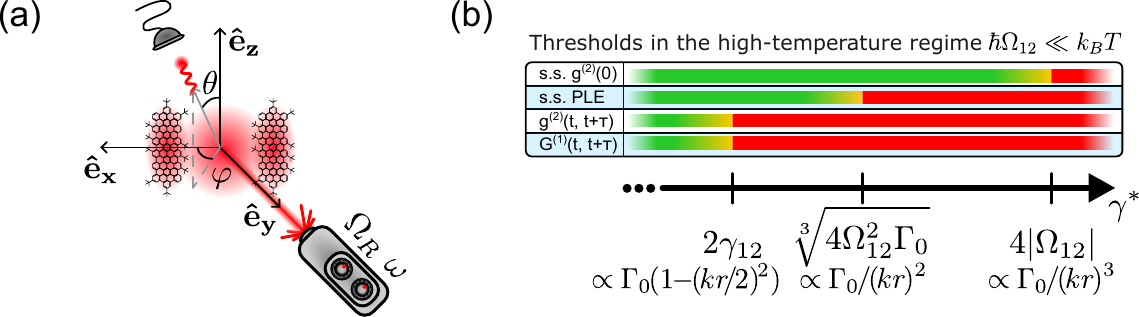}
\caption{\label{fig:schemaexp} (a) Schematic view of a typical experimental configuration: the two emitters are driven by a coherent laser field, and the emitted photons are collected and detected by a camera. The emitters are located on the $x$ axis, with their dipole moments oriented along $\mathbf{\hat{e}_z}$, corresponding to an 'H' configuration.
(b) Overview of the principal results of this work: a scale of the critical dephasing rates $\gamma^*_{\mathrm{lim}}$ above which the collective signatures disappear for the different observables considered. These thresholds are obtained in the high-temperature regime $\hbar|\Omega_{12}|/k_BT\ll1$, where the global master equation reduces to the local master equation.}
\end{figure*}

The downside of solid-state emitters is their strong interaction with the surrounding environment, which gives rise to inhomogeneous broadening and pure dephasing. The conventional view is that pure dephasing is detrimental to optical collective effects, which rely on phase coherence between emitters. This picture naturally arises within the local description of dephasing, which has been the predominant framework in the literature, where dephasing suppresses cooperative phenomena by destroying inter-emitter coherence. Consistent with this picture, superradiant and subradiant states in organic molecules have so far been predominantly observed at cryogenic temperatures, where pure dephasing is significantly reduced \cite{Sandoghdar2002,Trebbia2022,Lange2024}.
More recently, however, this conventional picture has been revisited through global descriptions of dephasing, revealing a richer interplay between the emitters and their environment \cite{shishkov2019relaxation,sanchez2026environment}. In this framework, the effect of dephasing depends on the temperature and does not necessarily reduce to a simple loss of coherence. Instead, it may induce transitions between collective states, redistributing population between superradiant and subradiant modes while preserving collective behavior \cite{vovcenko2021dephasing} as illustrated in Fig.~\ref{fig:intro} (b), which summarizes the crossover from the local to the global dephasing description.
Perovskite nanocrystals provide a particularly appealing platform to investigate the interplay between collective emission and dephasing. They can self-assemble into large one-, two-, or three-dimensional arrays with interparticle distances of only $d\sim10$--$30$ nm \cite{ruby2024influence,raino2018superfluorescence}. This structural organization has enabled the experimental observation of superradiant fluorescence bursts in such superlattices \cite{raino2018superfluorescence}. Remarkably, these bursts have even been observed at room temperature \cite{huang2022room,biliroglu2022room}, despite the strong pure dephasing expected under these conditions. These observations motivate a quantitative investigation of how environmental interactions modify experimentally accessible signatures of cooperative dynamics.
In this work, we address this question by quantitatively characterizing the influence of pure dephasing on collective emission. We first clarify the regimes of validity of global and local descriptions of dephasing and discuss how the relevant description depends on temperature. While the global approach reveals a richer dynamics, where dephasing can induce transitions between collective states, the local description~\cite{skinner1986pure,naesby2008influence,auffeves2009pure,Auffeves2010,cui2006emission} becomes the relevant framework in the high-temperature regime relevant to many solid-state experiments. We therefore focus on this regime and systematically investigate how increasing dephasing rates affect collective signatures through four complementary observables. This analysis reveals that the impact of dephasing is not universal: even within the local description, the critical dephasing rates at which collective signatures disappear strongly depend on the observable considered.
Concerning the light–matter interaction, we consider the minimal system of two coupled emitters ($N=2$) required to observe collective effects. This choice allows us to fully capture the system’s dynamics, both numerically and analytically, while keeping the Hilbert space dimension limited to $2^N=4$ [see Fig.~\ref{fig:intro}(a)]. Unlike previous works~\cite{celardo2014cooperative,shammah2017superradiance,kirton2017suppressing}, which often relied on simplified descriptions of light–matter interactions to access larger ensembles, either through nearest-neighbor approximations or Dicke-like models~\cite{dicke1954coherence}, we retain the full dipole–dipole interaction.
The remainder of the paper is organized as follows. In Section 2, we present the theoretical model and discuss the regimes of validity of the local and global master equations. Section 3 focuses on steady-state observables, while Section 4 addresses transient observables. For all observables, we vary $\gamma^*$ from negligible to dominant values and determine both analytically and numerically the critical dephasing rates beyond which collective signatures disappear [see Fig.~\ref{fig:schemaexp}(b)].
Although restricted to two emitters, our approach is not merely a toy model. Indeed, the dipole–dipole interaction in free space is extremely strong at short distances and decays rapidly with separation~\cite{agarwal2012quantum}. As a result, nearest-neighbor interactions often provide a good approximation for random systems such as hot vapors or ensembles of randomly distributed QDs~\cite{goetschy2011light,schilder2017homogenization,delpy2025anomalous,delpy2026non}. This work therefore contributes to a better understanding of the interplay between dephasing and dipole-dipole interactions across a broad range of physical platforms.

\section{Model}
In this paper we consider a pair of quantum emitters described by two-level systems (TLS). The state of each TLS can be written in the basis $\{|g_i\rangle, |e_i\rangle\}$, $i=1,2$, corresponding to the ground and excited states respectively~\cite{vivas2021two}. These emitters  exchange excitations with the free-space quantized electromagnetic modes and  interact with a  dephasing bath through a local phase-breaking interaction, as represented in Fig.~\ref{fig:intro}~(a). The total Hamiltonian of this system can be written as
\begin{equation}
H_{tot}=H_{mat}+H_{ph}+H_{B}+H_{mat-ph}+H_{mat-B},
\label{eq:Htot}
\end{equation}
where we expressed it as the sum of the matter (emitters) part, the photonic part, the unspecified dephasing bath part, the interaction between the emitters and the photons, and the interaction between the emitters and the dephasing bath.
A complete description of a pure state using Eq.~(\ref{eq:Htot}) is generally out of reach, as it would require an exact characterization of the infinitely many degrees of freedom of both the photonic and dephasing baths. Instead, we assume that (i) both baths are composed of harmonic oscillators, (ii) the emitters couple collectively to the same photonic bath $H_{ph}=\hbar \sum_{k}\omega_k a^\dagger_k a_k$, and  (iii) each emitter couples independently to its own dephasing bath $H_{B_i}=\hbar \sum_{k}\omega_k b^\dagger_{k,i} b_{k,i}$, such that $H_B=\sum_i H_{B_i}$. The properties of these dephasing baths and their temperature dependence are detailed in Appendix E. We model the interaction with the dephasing bath via a diagonal coupling, which implies no exchange of excitations between the emitter $i$ and the bath: $
H_{mat-B_i}=\hbar\sum_{k}\gamma_{k,i}\sigma_i^{ee}(b^\dagger_{k,i}+b_{k,i})$,
where $\sigma_i^{ee}=\sigma_i^\dagger\sigma_i$ and $\gamma_{k,i}$ is the coupling strength to the $k$-th bath mode.
The light-matter interaction is described by the standard Hamiltonian derived in the dipole and rotating-wave approximations:
$H_{mat-ph}=\hbar\sum_k\sum_i(g_k \sigma_i^\dagger a_k e^{i\mathbf{k}\cdot\mathbf{r_i}}+g_k^*a_k^\dagger \sigma_i e^{-i\mathbf{k}\cdot\mathbf{r_i}})$
~\cite{agarwal2012quantum,reitz2022cooperative},
where $g_k$ is the coupling strength between the emitters and the $k$-th photonic mode, and $\mathbf{r_i}$ denotes the position of emitter $i$.

\subsection{\label{sec_photon_coupling}Coupling to the photonic bath and light-mediated interactions}
To isolate the role of photon-mediated interactions, we first analyze the coupling to the common photonic bath in the absence of dephasing baths.
Following the usual procedure valid in the Born and Markov approximation~\cite{gardiner2004quantum,lambropoulos2007fundamentals,agarwal2012quantum,reitz2022cooperative}, we trace over the degrees of freedom of the photonic bath in order to obtain a master equation for the evolution of the density matrix of the TLSs degrees of freedom only. It writes:
\begin{equation}
    \frac{d\rho}{dt} = - \frac{i}{\hbar} [H, \rho] + \sum_{i,j = 1}^{2}\frac{\gamma_{ij}}{2}L_{\sigma_i, \sigma_j}[\rho] .
    \label{eq:Mastereq}
\end{equation}
In this expression, $\rho$ is a $4\times4$ density matrix that represents the mixed states of the two emitters interacting with the electromagnetic bath. It evolves in the space constructed from the pure state basis $\{|gg\rangle,|ge\rangle,|eg\rangle,|ee\rangle\}$ through two different terms written Eq.~(\ref{eq:Mastereq}). The first term corresponds to the Hamiltonian contribution, which includes the matter term written in Eq.~(\ref{eq:Htot}) and a dipole-dipole interaction term
\begin{equation}
    H = \underbrace{H_1 + H_2}_{H_{mat}} + \hbar\Omega_{12}(\sigma_1^\dagger\sigma_2 + \sigma_2^\dagger\sigma_1).
    \label{eq:Heff}
\end{equation}
In this expression $H_i/\hbar = (\omega_i-\omega)\sigma_i^\dagger\sigma_i + \frac{\Omega_{R_i}}{2}(\sigma_i + \sigma_i^\dagger)$ is the single emitter Hamiltonian expressed in terms  of $\omega_i$ the resonant frequency of the TLS, from which we define a detuning between emitters $\Delta = \omega_2 - \omega_1$. $\omega$ and $\Omega_{R_i}$ are respectively the frequency and the  Rabi frequency of a coherent driving laser field.  
Because we have eliminated the photonic degrees of freedom, the total Hamiltonian includes an additional term describing the coherent exchange of excitation through the  photonic bath between the two emitters, with a coupling amplitude $\Omega_{12}$. This dipole–dipole interaction couples the singly excited states $|eg\rangle$ and $|ge\rangle$, giving rise to the symmetric and antisymmetric combinations $|S\rangle = (|ge\rangle + |eg\rangle)/\sqrt{2}$ and $|A\rangle = (|eg\rangle - |ge\rangle)/\sqrt{2}$. Together with the ground state $|G\rangle = |gg\rangle$ and the doubly excited state $|E\rangle = |ee\rangle$, these form the eigenbasis of the system in the ideal case without detuning or external driving [see Fig.~\ref{fig:intro}~(a)].
The second term on the right-hand side of Eq.~(\ref{eq:Mastereq}) is expressed using the Lindblad superoperators $L_{\mathcal{O}_1,\mathcal{O}_2}[\rho] = 2\mathcal{O}_1\rho\mathcal{O}_2^\dagger - \{\mathcal{O}_2^\dagger\mathcal{O}_1, \rho\}$~\cite{stefanini2025lindblad}. It describes the irreversible loss of excitations into the shared electromagnetic vacuum. Collective effects arise from the double sum over distinct emitters $(j \neq i)$. The single-emitter radiative decay rate is $\gamma_{11} = \Gamma_0 = \gamma_{22}$, while $\gamma_{12} = \gamma_{21}$ represents photon exchange between emitters via dissipative free-space coupling.
For optical transitions, the thermal occupation of the photonic bath is negligible under experimentally relevant conditions, such that the electromagnetic environment can be effectively treated as being at zero temperature. Consequently, the photonic contribution to Eq.~(\ref{eq:Mastereq}) is independent of temperature and describes spontaneous emission without thermally induced absorption. A quantitative justification of this approximation is provided in Appendix~\ref{app:The baths}.

Both the coherent and dissipative couplings, $\Omega_{12}$ and  $\gamma_{12}$,  originate from  the free space dyadic electromagnetic Green's function  $\overline{\mathbf{G(\bf{r},\omega)}}$ renormalized by  the Debye-Waller/Franck-Condon factor $\alpha$:
\begin{eqnarray}
    \overline{\mathbf{G}(\bf{r},\omega)} =&&-\frac{3}{4}\alpha\Gamma_0\frac{e^{ikr}}{kr}\biggl[\bm{I}_3\left(1 + \frac{i}{kr} - \frac{1}{(kr)^2}\right) \nonumber \\
    &&+ \mathbf{\hat{r}\hat{r}^T}\left(-1 - \frac{3i}{kr} + \frac{3}{(kr)^2}\right)\biggr].
        \label{eq:Green}
\end{eqnarray}
They can be computed through the formulas: $\Omega_{12}=\text{Re}\left[G\right]$ and $\gamma_{12} = -2\text{Im}\left[G\right]$, with $G = \mathbf{\hat{d}^T_1  \overline{\mathbf{G}} \hat{d}_2}$.
In Eq.~(\ref{eq:Green})  $k = n(\omega_1 + \omega_2)/2c$ is the average wavevector amplitude  in a medium of refractive index $n$, $\Gamma_0$ the radiative decay rate in this medium, $c$ the speed of light in vacuum, $r$ the distance between the emitters, $\bm{I}_3$ the $3\times 3$ identity matrix. We denote all unit vectors with a hat: $\mathbf{\hat{r}}$ the unit vector pointing from one emitter towards the other and $\mathbf{\hat{d}_i}$ the unit vector pointing in the direction of the dipole moment of emitter $i$. 

The Debye-Waller/Franck-Condon factor $\alpha$ quantifies the fraction of emission from the excited state decaying directly to the ground state, commonly known in solid-state emitters as the zero-phonon line (ZPL), as opposed to red-shifted emission from the excited state decaying into vibrational energy levels above the ground state [see Fig.~\ref{fig:intro}~(a)].
Because these two processes are spectrally distinct, the total electric field radiated by the emitters naturally separates into a ZPL contribution and a vibronic (red-shifted) contribution:
\begin{align}
    \mathbf{E}_{tot} = \mathbf{E}_{\text{ZPL}} + \mathbf{E}_{vib},
\end{align}
which can be measured separately using frequency filters in quantum optics experiments~\cite{Costanza2021Indistinguishable, Sandoghdar2010Indistinguishable}.
In this work, we describe completely the degrees of freedom involved in the ZPL decay which occurs at a rate $\alpha\Gamma_0$. This treatment enables us to incorporate the dipole–dipole coupling between the two emitters and to study in detail how collective effects compete with pure dephasing in determining the statistics of $\mathbf{E}_{\text{ZPL}}$.
The remaining fraction of the decay, which proceeds through vibronic states at a rate $(1-\alpha)\Gamma_0$, is formally treated as non-radiative. 
Indeed, the vibronic states involved have extremely short lifetimes, typically on the order of a picosecond or less \cite{Quistrebert2026}. As a consequence, the corresponding red-shifted emission from different emitters does not interfere.
This minimal treatment of the vibronic emission still permits a theoretical description of measurements of the total intensity emitted in the red-shifted vibronic replica, as detailed in Section \ref{sec_EP}.
We note that an explicit analytical description of the field $\mathbf{E}_{vib}$ would require modeling the emitters as (at least) three-level systems as done recently in~\cite{JuanDelgado2025}.
In organic molecules typically used for quantum optics, $\alpha \approx 0.3$. Their emission usually lies in the red–visible region of the spectrum ($\lambda \approx 600-800$nm), and exhibits lifetimes on the order of a few nanoseconds ($\Gamma_0/2\pi \approx$ few $10$s of MHz). Pairs of molecules have been reported with a coupling energy of a few $10$s of $\hbar\Gamma_0$, corresponding to distances of a few $10$s of nm, and with dissipative coupling rates approaching the maximum value $\gamma_{12} \approx \alpha \Gamma_0$ ~\cite{Trebbia2022, Lange2024}. Accordingly, in all figures presented in this work, we use experimentally realistic parameters: $\Omega_{12} = 20.0\Gamma_0$ and $\gamma_{12} = 0.3\Gamma_0$
\subsection{\label{sec_deph_coupling}Description of pure dephasing in dipole-coupled emitters}
The derivation of the dephasing master equation requires particular care in the presence of light-mediated dipole-dipole interactions. Indeed, these interactions modify the eigenstates of the coupled emitters. Neglecting them in the system Hamiltonian used to derive the incoherent part of the master equation  leads to the local description, whereas retaining them yields the global one. We briefly present both approaches below before focusing on the local description, which constitutes the main framework of this work.
\subsubsection{Local master equation}
In the local approach, the dipole-dipole interaction is neglected when tracing over the dephasing baths. Consequently, the interaction-picture operators appearing in the dephasing interaction Hamiltonian $H_{mat-B_i}$ satisfy $\sigma_i^{ee}(t)\simeq\sigma_i^{ee}$, since they commute with the bare system Hamiltonian. Assuming that the photonic bath and the two dephasing baths are uncorrelated, each bath can then be traced out independently, yielding the master equation:
\begin{equation}
    \frac{d\rho}{dt} = - \frac{i}{\hbar} [H, \rho] + \sum_{i,j = 1}^{2}\frac{\gamma_{ij}}{2}L_{\sigma_i, \sigma_j}[\rho] + \sum_{i = 1}^2 \frac{\gamma^*(T)}{2}L_{\sigma_i^{ee}}[\rho].
    \label{eq:Mastereq_dephasing_local}
\end{equation}
Compared to Eq.~\eqref{eq:Mastereq}, the master equation contains an additional local dephasing contribution describing the coupling of each emitter $i$ to its own dephasing bath through the superoperators $L_{\sigma_i^{ee}}[\rho]\equiv L_{\sigma_i^{ee},\sigma_i^{ee}}[\rho]$.
Because this term originates from a diagonal interaction with the bath, implying no exchange of excitations, it conserves the TLS populations while damping the coherences at the pure dephasing rate $\gamma^*(T)$~\cite{Auffeves2010, Shlesinger2019}. In the local description, the influence of temperature is entirely captured by the value of $\gamma^*(T)$, whose temperature dependence depends on the microscopic model used to describe the bath. Under the Markov approximation, for identical dephasing baths at the same temperature $T$, we obtain
\begin{equation}
    \gamma^*(T)/2 = \pi\int_0^{+\infty} d\omega J(\omega)  \delta(\omega) [2 n_{th}(\omega) + 1],
\end{equation}
where $n_{th}(\omega) = \left[e^{\hbar \omega/k_B T} - 1\right]^{-1}$ is the Bose-Einstein occupation number of the bath modes and $J(\omega)~=~\sum_k \lvert \gamma_k \rvert^2\delta(\omega - \omega_k)$ is the spectral density. We assume that the interaction with the dephasing bath can be described with an Ohmic spectral density $J(\omega)~=~C\omega e^{-\omega/\omega_c}$, where $\omega_c$ is a high-frequency cutoff that depends on the specific properties of the bath and $C$ is a dimensionless parameter quantifying the strength of the interaction with the bath. In this case, the dephasing rate reduces to
\begin{equation}
    \gamma^*(T) = 4\pi C k_B T/\hbar,
\end{equation}
where the linear temperature dependence is reminiscent of electron-phonon coupling at high temperatures. In the following, we omit the explicit temperature dependence and simply write $\gamma^* \equiv \gamma^*(T)$.
\subsubsection{Global master equation}
The local master equation provides a simple and widely used description of pure dephasing in interacting quantum systems. It is exact when the system Hamiltonian does not contain interaction terms between the emitters, as in the Dicke model, and has therefore been extensively employed in the literature to describe collective emission phenomena~\cite{celardo2014cooperative,shammah2017superradiance,kirton2017suppressing}. However, in the presence of finite dipole-dipole interactions, the eigenstates of the coupled system are modified [see Fig.~\ref{fig:intro}~(a)], and the validity of the local approach requires further consideration. 
 In particular, when the energy splitting induced by the dipole-dipole interaction becomes comparable to the characteristic dissipative timescales, neglecting this interaction in the derivation of the dephasing contribution may lead to thermodynamic inconsistencies~\cite{Levy_2014,shishkov2019relaxation}. In this regime, the system operators associated with the dephasing interaction can no longer be assumed to be time-independent, since $[H,\sigma_i^{ee}]\neq0$. In the Heisenberg-picture representation, they evolve as
\begin{align}
	\sigma_i^{ee}(t) &= e^{iH t/\hbar}\, \sigma_i^{ee}\, e^{-iH t/\hbar}, \notag \\
    &=\sum_{k,k'}  e^{i(\omega_k - \omega_{k'}) t}\, \langle k|\sigma_i^{ee} |k'\rangle\, |k \rangle \langle k'|,
    \label{eq:op_heisenberg}
\end{align}
where the vectors $\{|k\rangle\}$ are eigenstates of $H$ with eigenvalues $\{\hbar \omega_k\}$. In the absence of an external driving field and detuning ($\Omega_R = \Delta = 0$), Eq.~\eqref{eq:op_heisenberg} reduces to
\begin{equation}
    \sigma_{1/2}^{ee}(t) =  \sigma_{\text{D}} \pm \frac{1}{2}\left( e^{2i\Omega_{12}t} \sigma_\text{SA} +  e^{-2i\Omega_{12}t} \sigma_\text{AS}\right),
    \label{eq:op_heisenberg_collective}
\end{equation}
where $\sigma_\text{D} = | E \rangle \langle E | + \left(| S \rangle \langle S | + | A \rangle \langle A |\right)/2$, $\sigma_\text{AS} = |A \rangle \langle S |$ and $\sigma_\text{SA} = |S \rangle \langle A |$. Following the standard Davies procedure \cite{shishkov2019relaxation,Davies1974,Davies1976}, the global master equation is then obtained by performing a full secular approximation, which is valid as long as $\lvert \Omega_{12} \rvert/\Gamma_0 \gg 1$. The final equation reads
\begin{align}
    \frac{d\rho}{dt} &= - \frac{i}{\hbar} [H, \rho] + \sum_{i,j = 1}^{2}\frac{\gamma_{ij}}{2}L_{\sigma_i, \sigma_j}[\rho]  + \gamma^*(T) L_{\sigma_\text{D}}[\rho]  \notag \\
    &+ \Gamma_{+-}(T) L_{\sigma_\text{AS}}[\rho] + \Gamma_{-+}(T) L_{\sigma_\text{SA}}[\rho],
    \label{eq:Master_eq_global}
\end{align}
where $\Gamma_{+-}(T)$ and $\Gamma_{-+}(T)$ are the temperature-dependent transition rates between the collective states $|S\rangle$ and $|A\rangle$ represented in  Fig.~\ref{fig:intro}.
As for the local dephasing rate, explicit expressions for the transition rates can be obtained once a microscopic model for the bath is specified. Assuming an Ohmic spectral density, one finds:
\begin{align}
	&\Gamma_{+-}(T) = \gamma^*(T) \frac{\hbar \Omega_{12}}{k_B T} [n_{th}(2\Omega_{12}) - n_{th}(-2\Omega_{12}) + 1],\label{eq:transition_rate_SA}\\
	&\Gamma_{-+}(T) = \gamma^*(T) \frac{\hbar \Omega_{12}}{k_B T} [n_{th}(2\Omega_{12}) - n_{th}(-2\Omega_{12}) - 1].\label{eq:transition_rate_AS}
\end{align}
From Eqs.~\eqref{eq:transition_rate_SA} and \eqref{eq:transition_rate_AS}, two distinct temperature regimes naturally emerge. In the low-temperature regime, $\hbar |\Omega_{12}|/k_B T \gg 1$, thermally induced upward transitions are negligible, so that population transfer occurs only from the higher-energy collective state to the lower-energy one. In contrast, in the high-temperature regime, $\hbar |\Omega_{12}|/k_B T \ll 1$, upward and downward transitions become nearly equally probable, leading to $\Gamma_{-+}(T)\simeq\Gamma_{+-}(T)\simeq\gamma^*(T)$. Consequently, in the limit $\hbar |\Omega_{12}|/k_B T\rightarrow0$, the global master equation~\eqref{eq:Master_eq_global} continuously reduces to the local master equation~\eqref{eq:Mastereq_dephasing_local}, a crossover summarized schematically in Fig.~\ref{fig:intro}(b).  More generally, at arbitrary temperature, the global master equation can be rewritten as the local master equation supplemented by temperature-dependent correction terms accounting for population transfer between collective states.
For notational simplicity, we omit the explicit temperature dependence of $\gamma^*$, $\Gamma_{+-}$ and $\Gamma_{-+}$ in the following expression, while keeping in mind that these quantities remain temperature dependent:
\begin{align}
\label{eq:global_fct_local}
\frac{d\rho}{dt} &= - \frac{i}{\hbar} [H, \rho] + \sum_{i,j = 1}^{2}\frac{\gamma_{ij}}{2}L_{\sigma_i, \sigma_j}[\rho] +\sum_{i = 1}^2 \frac{\gamma^*}{2}L_{\sigma_i^{ee}}[\rho] \notag \\
    &+ \left(\Gamma_{+-} - \gamma^* \right) L_{\sigma_\text{AS}}[\rho] + \left(\Gamma_{-+} - \gamma^*\right) L_{\sigma_\text{SA}}[\rho].
\end{align}
This decomposition highlights the fact that the difference between the local and global equations becomes increasingly significant as the ratio $\hbar \lvert \Omega_{12} \rvert /k_B T$ grows larger.

For experimentally relevant parameters, this ratio is typically much smaller than unity. For instance, at $T=1\,\mathrm{K}$, we have $k_B T \approx 8.6\times10^{-5}\,\mathrm{eV}$, which can be compared with typical values of $\hbar\Omega_{12}$ for emitter systems, lying in the $\mu\mathrm{eV}$ range as reported in \cite{Trebbia2022,Lange2024}. Thus, the corrections associated with the local approximation are expected to be negligible under typical experimental conditions. In the remainder of this paper, we therefore focus on the regime $\hbar|\Omega_{12}|\ll k_B T$ and use the local master equation to describe dephasing. We refer the reader to Appendix~\ref{app:The baths} for a detailed analysis of the complementary low-temperature regime, including a comparison of the global and local master equations and a broader discussion of the validity of the local approximation.
\begin{figure*}
\includegraphics[width=\textwidth]{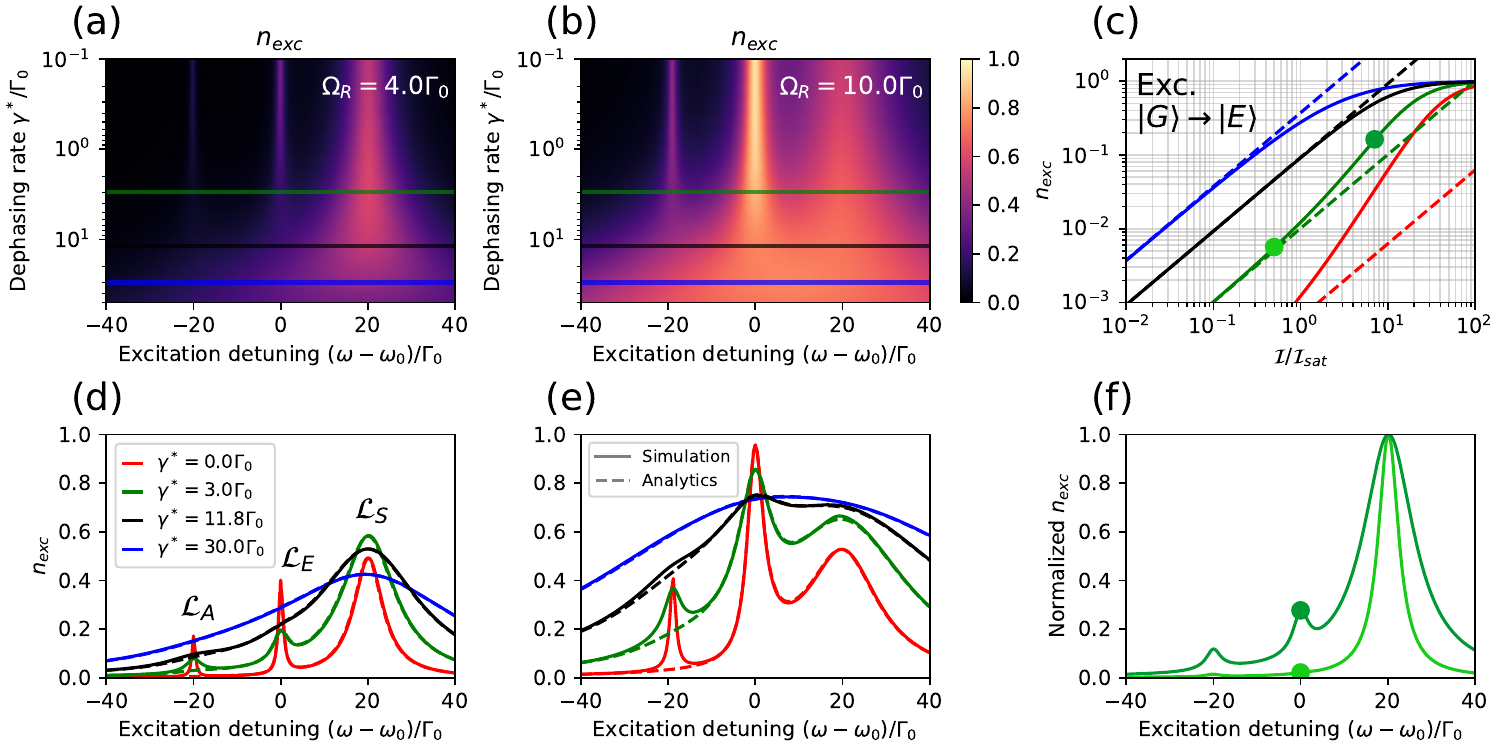}
\caption{\label{fig:ple/main} Dependence of the excitation spectra on the excitation intensity $\mathcal{I}$ and dephasing rate $\gamma^*$. Excitation spectra as a function of the dephasing rate for Rabi frequencies (a) $\Omega_R = 4.0\Gamma_0$ and (b) $\Omega_R = 10.0\Gamma_0$.  Saturation curves at $\omega = \omega_0$ (c) for $\gamma^* = 0$ (red), $\gamma^* = 3.0\Gamma_0$ (green), $\gamma^* = 30.0\Gamma_0$ (blue) and $\gamma^*_{lim} \approx 11.8\Gamma_0$ (black, see eq. \ref{eq:gamma^*_lim_saturation}). Dashed lines show the analytical expression's first order series expansion in $(\Omega_R/\Gamma_0)^2$ (see Appendix~\ref{app:steady_state_saturation}). Plain lines (d) and (e): cuts of (a) and (b) respectively. Dashed lines show the exact analytical expression with $\Delta = 0$ and all other parameters kept identical.  (f) Normalized excitation spectra for $\gamma^* = 3.0\Gamma_0$, $\mathcal{I}/\mathcal{I}_{sat} = 0.5$ (light green) and $7.0$ (dark green). The dots correspond to the dots in (c). In all panels, simulations are performed with a detuning between emitters of $\Delta = 5 \Gamma_0$, while analytical results are obtained for $\Delta=0$. The emitters are placed $r \approx 0.035\lambda$ apart, in an 'H' configuration, with a Debye-Waller/Franck-Condon factor $\alpha = 0.3$, yielding a coupling strength $\Omega_{12} = 20\Gamma_0$ and a dissipative coupling rate $\gamma_{12} = 0.3\Gamma_0$.}
\end{figure*}

\section{Steady-state observables}
In this section, we investigate two experimentally accessible steady-state observables within the local master-equation framework introduced above. The first is the excited-state population of the TLS as a function of the incident laser frequency $\omega$, which has been shown to reproduce well the steady-state photoluminescence excitation spectra of pairs of coupled molecules when only the red-shifted emission of $\mathbf{E}_{vib}$ into phonon replicas is collected~\cite{Sandoghdar2002} (see Appendix~\ref{app:redshifted_intensity}). The second is the zero-delay second-order correlation function of the zero-phonon-line field, $g^{(2)}(0)$. In both cases, the two emitters are driven by a monochromatic electromagnetic field. Since, within the local description, the temperature dependence of the dephasing environment is fully captured by the dephasing rate $\gamma^*$, we use this parameter as a control variable to quantify the impact of pure dephasing on collective signatures. Our goal is to determine how these observables evolve with increasing dephasing and to identify the critical values of $\gamma^*$ beyond which collective effects are no longer observable.
To access the steady state of our system, we write the master equation given in Eq.~(\ref{eq:Mastereq_dephasing_local}) in a compact form by introducing the Liouvillian superoperator $\mathcal{L}$:
\begin{equation}
    \frac{d\rho}{dt} = \mathcal{L}[\rho].
    \label{eq:Liouville}
\end{equation}
The steady-state density matrix $\rho_{ss}$ is then found by solving
$\mathcal{L}[\rho_{ss}] = 0$ numerically and analytically.
We use the symbolic computation library SymPy ~\cite{Meurer2017} to obtain the analytical expression of $\rho_{ss}$  in the (slightly restrictive) case of two identical emitters, with no detuning between emitters ($\Delta = 0$) and identical driving strengths ($\Omega_{R_1} = \Omega_{R_2} = \Omega_R$). The analytical expression can be found in~\cite{lumin_analytics}.

\subsection{\label{sec_EP}Excited-state population}
Our first observable of interest is the number of excitations in the emitters denoted $n_{exc}$. 
It is defined as the sum of the excited state populations of each emitter and can be expressed in terms of the coupled density matrix elements: $n_{exc} = 2\rho_{ee,ee} + \rho_{eg,eg} + \rho_{ge,ge}$, where we use the notation $\rho_{ab,cd} = \langle a,b|\rho|c,d\rangle$. 
In Fig.~\ref{fig:ple/main}~(a) we present a 2D map of the excited-state population $n_{exc}$, obtained from the numerical resolution of Eq.~(\ref{eq:Mastereq_dephasing_local}). The map is shown as a function of both the detuning between the laser and emitter frequencies $\omega - \omega_0$ with $\omega_0 = (\omega_1 + \omega_2)/2$, and the pure dephasing rate, for a relatively large Rabi frequency $\Omega_R = 4\Gamma_0$.
For clarity, panel (d) shows horizontal cuts of (a) at fixed dephasing rates, with each value indicated by a different color.
At low dephasing, three distinct peaks appear, noted $\mathcal{L}_A$, $\mathcal{L}_E$ and $\mathcal{L}_S$, centered around detunings $\omega - \omega_0 = -\Omega_{12}$, $0$ and $+\Omega_{12}$ respectively.
These correspond to transitions from the ground state $|G\rangle$ to the antisymmetric $|A\rangle$, doubly excited $|E\rangle$, and symmetric $|S\rangle$ states, in close agreement with experimental observations~\cite{Sandoghdar2002, Trebbia2022, Lange2024}.
We provide analytical expressions of these peaks extracted from~\cite{lumin_analytics} and simplified in various regimes in Appendix~\ref{app:steady_state_lorentzians}.
The relatively large $\Omega_R$ used here is essential to make the central two-photon peak clearly visible.
As the dephasing rate increases, these peaks broaden and merge into one broad band while their positions remain fixed since dephasing affects only the dissipative part of the dynamics.\\
In Fig.~\ref{fig:ple/main}(b) and (e), we show similar results for a stronger drive, $\Omega_R = 10\Gamma_0$, large enough to nearly saturate the two-photon transition. This transition saturates at $n_{exc} = 1$, corresponding to half the maximum possible excitation of two two-level systems. Power broadening is clearly observed in all peaks, with the $\mathcal{L}_A$ peak being the least affected owing to its weaker coupling to light.
In panels (d) and (e), we compare the numerical results (solid lines) with our analytical expression for $n_{exc}$ (dashed lines), finding excellent agreement across all dephasing regimes, except for $\mathcal{L}_A$. This discrepancy is expected, since our analytical model assumes zero detuning, for which the subradiant mode is perfectly dark.\\
In  particular, the perfect agreement between numerics and analytics for the central peak $\mathcal{L}_E$ is important. It permits us to model practical experiments on molecules, where coupled pairs of emitters are found by searching for spectral lines whose amplitude scale quadratically with the excitation power ($\propto \Omega_R^4$).
These lines correspond to the two-photon transition from the ground state to the doubly excited state $|G\rangle \rightarrow |E\rangle$. Because dephasing destroys coherent coupling, we expect this quadratic scaling to vanish above a certain dephasing threshold $\gamma^*_{lim}$, making the detection of coupled pairs increasingly difficult.
In Fig.~\ref{fig:ple/main}(c), we show simulated saturation curves of the central resonance, $n_{exc}(\omega = \omega_0)$, as a function of excitation intensity. Solid lines correspond to the same dephasing values as in (d) and (e), while dashed lines represent the linear approximations derived from our analytical model.
For small $\gamma^*$ (red and green curves), we observe a clear superlinear growth before saturation, characteristic of a two-photon process. In contrast, for large $\gamma^*$ (blue curve), the superlinear regime disappears, leaving only a linear dependence before saturation.\\
\begin{figure*}
\includegraphics[width=\textwidth]{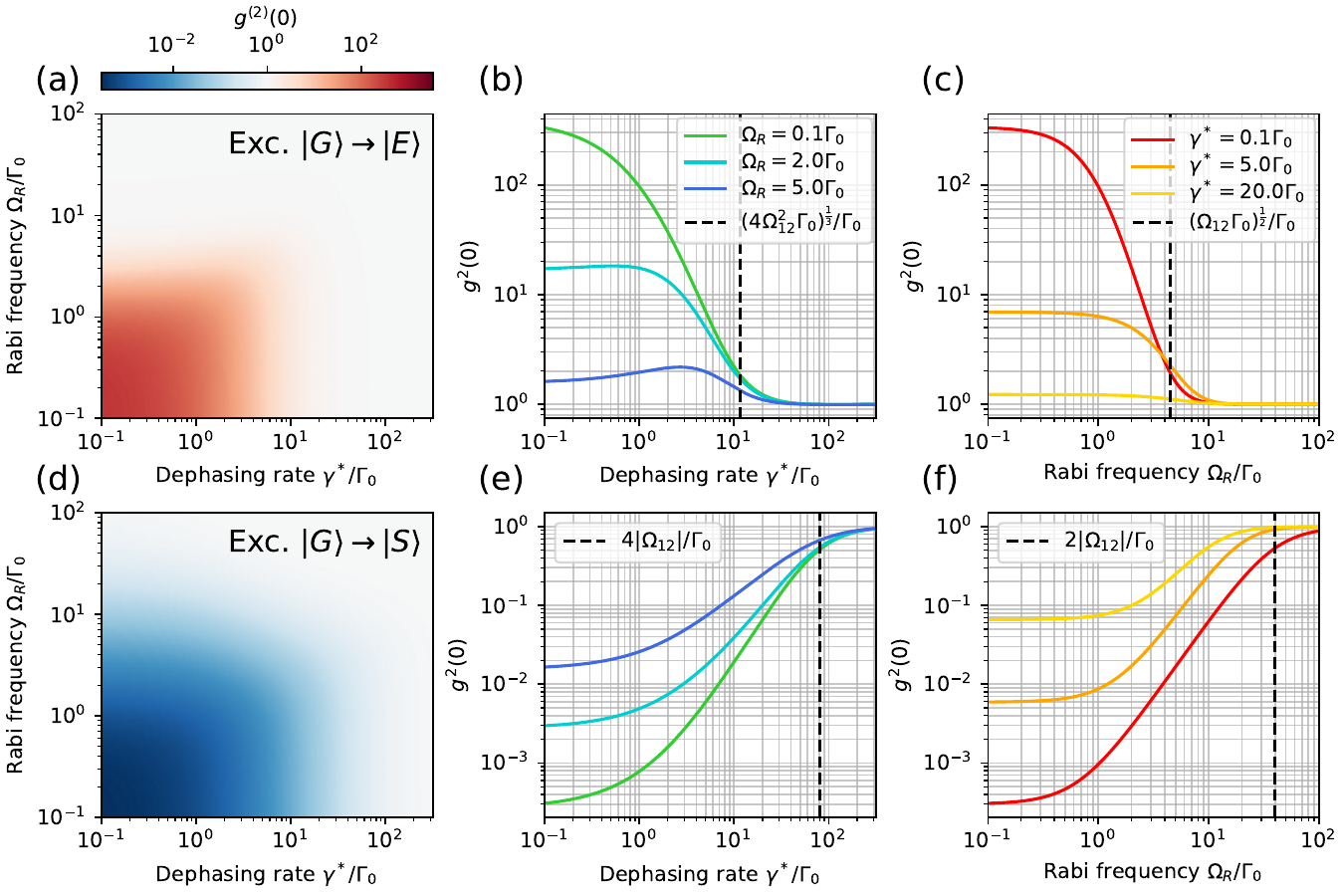}
\caption{\label{fig:steady_state_g2/main} (a) Second-order correlation at zero delay $g^{(2)}(0)$ as a function of the Rabi frequency $\Omega_R$ and dephasing rate $\gamma^*$, with the excitation tuned to the two-photon transition $\omega=\omega_0$. For large $\Omega_R$ or $\gamma^*$,  $g^{(2)}(0) \rightarrow 1$, indicating poissonian statistics. (b) Horizontal cuts of (a) for different Rabi frequencies $\Omega_R = 0.1$ (green), $2.0$ (light blue) and $5.0 \Gamma_0$ (dark blue). (c) Vertical cuts of (a) for different dephasing rates $\gamma^* = 0.1$ (red), $5.0$ (orange), $20.0 \Gamma_0$ (yellow). (d), (e) and (f): same as (a), (b) and (c) with the excitation tuned to the $|G\rangle \rightarrow |S\rangle$ transition. The emitters are identical ($\Delta = 0$), placed $r \approx 0.035\lambda$ apart, in an 'H' configuration, with a Debye-Waller/Franck-Condon factor $\alpha = 0.3$, yielding a coupling strength $\Omega_{12} = 20\Gamma_0$ and a dissipative coupling rate $\gamma_{12} = 0.3\Gamma_0$.}
\end{figure*}
Using our analytical expression, we obtain an upper bound on the dephasing rate $\gamma^*_{lim}$ (black curve) above which the faster-than-linear scaling disappears completely. In the limit $\Omega_{12} \gg \Gamma_0$, it writes:
\begin{equation}
    \gamma^*_{lim} \approx \left(4\Omega_{12}^2\Gamma_0\right)^\frac{1}{3}.
    \label{eq:gamma^*_lim_saturation}
\end{equation}
Remarkably, $\gamma^*_{lim}$ depends only on the coupling strength $\Omega_{12}$, which can be very large in solid-state emitters thanks to their strong dipole moments and sub-wavelength separations. This suggests that identifying coupled emitter pairs remains feasible even when their emission linewidth is not lifetime-limited, facilitating exploratory studies on new emitter platforms.
Finally, even for $\gamma^*=0$, the low-intensity scaling deviates slightly from an ideal quadratic law because the signal at $\omega=\omega_0$ still contains a small linear contribution from neighboring resonances—primarily the tail of $\mathcal{L}_S$, as visible in panel (f).

\subsection{\label{section_g2}Second-order correlations at zero delay $g^{(2)}(0)$}
Let us now turn to the study of the second-order correlations of the radiated electric field $\mathbf{E}_{\text{ZPL}}$ in the steady state and at zero delay: $g^{(2)}(0)$. This quantity gives valuable insight on the statistics of the photon emission. In the particular case of two emitters, it is known to be sensitive to the dipole-dipole coupling between them \cite{vivas2021two}. In this section, we investigate how the dephasing rate influences  $g^{(2)}(0)$, and we identify a theoretical threshold value of $\gamma^*$ above which the signatures of the coherent coupling are lost. 
This effect is reminiscent of the decorrelating effect of the Rabi frequency $\Omega_R$ on the two-body density matrix reported in previous works \cite{juan2024tailoring,goncalves2025driven}.
To compute $g^{(2)}(0)$ we first recall that, in the far-field and near resonance, the positive-frequency component of $\mathbf{E}_{\text{ZPL}}$ at position $\mathbf{R}$ is \cite{CARMICHAEL2000417}:
\begin{eqnarray}
    \label{eq:radiated_field}
    \mathbf{E}_{\text{ZPL}}^{(+)}\left(\mathbf{R}, t+\frac{n R}{c}\right) =&& -\frac{k^2d}{4\pi\varepsilon_0}\\
    &&\times\sum_{i=1}^2\frac{(\mathbf{\hat{d}_i}\times \mathbf{\hat{R}_i})\times\mathbf{\hat{R}_i}}{R_i}e^{i{\color{red}k}(R_i-R)}\sigma_i(t),\nonumber
\end{eqnarray}
with $\mathbf{R}_i = \mathbf{R}-\mathbf{r}_i$ and $k = n(\omega_1 + \omega_2)/2c$. We take $\mathbf{\hat{d}_1} = \mathbf{\hat{d}_2} = \mathbf{\hat{e}_z}$ as illustrated in Fig.~\ref{fig:schemaexp} (a), and $\mathbf{R}$ in the $\text{Oxy}$ plane with $R \gg r$. Under these conditions, and using $d^2=3\pi\varepsilon_0\hbar \alpha\Gamma_0/k^3$, Eq.~\ref{eq:radiated_field} simplifies to:
\begin{equation}
    \mathbf{E}_{\text{ZPL}}^{(+)}\left(\mathbf{R}, t+\frac{ nR}{c}\right) = -\mathbf{\hat{e}_z}\frac{1}{4R}\sqrt{\frac{3\alpha\hbar k\Gamma_0}{\pi\varepsilon_0}}\sum_{i=1}^2e^{-ik\mathbf{\hat{R}}\cdot\mathbf{r_i}}\sigma_i(t).
\end{equation}
Since only the relative phases and amplitudes matter for the normalized correlation function, we omit the overall prefactor (constant for fixed $R$) and focus instead on the dimensionless quantity:
\begin{equation}
    D = \sum_{i=1}^2\frac{e^{-i{\color{red}k}\mathbf{\hat{R}}\cdot\mathbf{r_i}}}{\sqrt{2}}\sigma_i(t) = \frac{e^{ik\mathbf{\hat{R}}\cdot\mathbf{r}/2}}{\sqrt{2}}\sigma_1(t) + \frac{e^{-ik\mathbf{\hat{R}}\cdot\mathbf{r}/2}}{\sqrt{2}}\sigma_2(t).
    \label{eq:Dvector}
\end{equation}
where $\mathbf{r}=\mathbf{r_2}-\mathbf{r_1}$.
With this definition, the second-order correlations at zero delay $g^{(2)}(0)$ can be computed as:
\begin{equation}
    g^{(2)}(0) = \frac{\langle D^\dagger D^\dagger D D \rangle}{\langle D^\dagger D\rangle^2}.
    \label{eq:G2_andD}
\end{equation}
As seen from Eq. \ref{eq:Dvector}, the operator that represents the emission of light by two emitters  explicitly depends on the observation direction. Consequently, the correlation function $g^{(2)}(0)$ also exhibits a directional dependence studied in previous works \cite{masson2020many}.
For simplicity, in this section we restrict the discussion to the case where the detection is perpendicular to the inter-emitter axis, $\mathbf{R} = R\mathbf{\hat{e}_y}$ (see Fig. \ref{fig:schemaexp} (a)), such that $D = \frac{\sigma_1 + \sigma_2}{\sqrt{2}}$. Injecting this expression into Eq. \ref{eq:G2_andD} leads to:
\begin{equation}
    g^{(2)}(0) = \frac{4\rho_{ee,ee}}{(2\rho_{ee,ee} + \rho_{eg,eg} + \rho_{ge,ge} + \rho_{eg,ge} + \rho_{ge,eg})^2}.
    \label{eq:G2_value}
\end{equation}
Equation \ref{eq:G2_value} expresses $g^{(2)}(0)$ in terms of the different elements of the steady-state two-body density matrix. 
It provides the basis for analyzing how photon correlations depend on both the system parameters and the driving conditions.
To illustrate these dependencies, Fig.~\ref{fig:steady_state_g2/main} presents numerical evaluations of $g^{(2)}(0)$ in the steady state for different values of the Rabi frequency and the pure dephasing rate.
Two excitation frequencies are considered:  (a) the two-photon transition  $|G\rangle \rightarrow |E\rangle$ and (d)  the $|G\rangle \rightarrow |S\rangle$ transition. 
For clarity, horizontal cuts of panels (a) and (d) are shown as solid lines in panels (b) and (e), respectively.
The different colors  correspond to various  Rabi frequencies. Similarly, vertical cuts of panels (a) and (d) are displayed in panels (c) and (f), with colors indicating different dephasing rates.\\
In Fig.~\ref{fig:steady_state_g2/main}(a,d), two distinct regimes can be identified.
In the first regime, corresponding to small values of both $\Omega_R$ and $\gamma^*$ collective effects play a dominant role in shaping the photon statistics of the emitted light.
When the driving field is tuned to the two-photon $|G\rangle\rightarrow|E\rangle$ transition [Fig. \ref{fig:steady_state_g2/main} (a)], pronounced photon bunching is observed with $g^{(2)}(0) \gg 1$.
However, this bunching does not imply that photon-pair emission dominates the dynamics as the population of the doubly excited state $|E\rangle$ remains significantly smaller than that of the superradiant state $|S\rangle$.
Conversely, when the excitation is resonant with the $|G\rangle \rightarrow |S\rangle$ transition, the doubly-excited state $|E\rangle$ remains essentially unpopulated.
In this case, the two emitters behave collectively as a single-photon source, and pronounced antibunching is observed with $g^{(2)}(0) \ll 1$.

The second regime corresponds to large values of either $\Omega_R$ or $\gamma^*$  and is characterized by $g^{(2)}(0) \approx 1$. Within this parameter range, the two emitters radiate independently, and the photon statistics becomes Poissonian.
Although $\Omega_R$ and $\gamma^*$ seem to play similar roles at first sight, the physical mechanisms leading to $g^{(2)}(0) \approx 1$ differ markedly in the two cases.
When the Rabi frequency is large and the dephasing rate remains small, the emitters become saturated. As reported in \cite{juan2024tailoring}, saturation tends to destroy quantum correlations, and the two-body density matrix $\rho$  becomes factorizable, i.e.  $\rho = \rho_1\otimes\rho_2$, where $\rho_1$ and $\rho_2$ denote the single-emitter density matrices. In this high-excitation regime, and to lowest order in $1/\Omega_R$, the coherences vanish and the populations equalize:
$\rho_1=\rho_2\simeq\begin{pmatrix}
1/2 & 0 \\
0 & 1/2 
\end{pmatrix}$.
Conversely, when the dephasing rate $\gamma^*$ is large and $\Omega_R$ is  small, the entanglement is suppressed by pure dephasing. The two-body density matrix again factorizes as $\rho = \rho_1\otimes\rho_2$, but the system now remains mostly in its ground state
$\rho_1=\rho_2\simeq\begin{pmatrix}
1 & 0 \\
0 & 0 
\end{pmatrix}$,
 to the lowest order in $1/\gamma^*$.
This separability condition can be formulated quantitatively for each component of the two-body density matrix. In Appendix~\ref{app:g2_0_thresholds}, we apply it to the doubly excited population to derive analytical thresholds for $\gamma^*$ and $\Omega_R$ beyond which entanglement vanishes.
These thresholds are summarized in Table \ref{tab:g2_criteria} and plotted as black dashed lines in panels (b), (c), (e) and (f) of Fig.~\ref{fig:steady_state_g2/main}. They also correspond to $g^{(2)}(0) \approx 2$ for $\omega=\omega_0$ and $g^{(2)}(0) \approx \frac{1}{2}$ for $\omega-\omega_0 = \Omega_{12}$.
\begin{table}[htbp]
\caption{\label{tab:g2_criteria} Thresholds}
\begin{ruledtabular}
\begin{tabular}{lcr}
    Excitation: & $\omega = \omega_0$ & $\omega-\omega_0 = \Omega_{12}$ \\
    \colrule
    $\gamma^*_{lim}$ & $(4\Omega_{12}^2\Gamma_0)^\frac{1}{3}$ & $4|\Omega_{12}|$ \\
    $\Omega_{R_{lim}}$ & $(|\Omega_{12}|\Gamma_0)^\frac{1}{2}$ & $2|\Omega_{12}|$ \\
\end{tabular}
\end{ruledtabular}
\end{table}\\
To interpret those values we refer to the excitation spectra shown in Fig.~\ref{fig:ple/main}(d–e). For small $\gamma^*$ and $\Omega_R$ the spectral lines associated to $|G\rangle \rightarrow |S\rangle$  and $|G\rangle \rightarrow |E\rangle$  are well resolved and separated by $\Delta \omega=\Omega_{12}$. In this regime, it is possible to selectively excite either the single-photon transition (leading to antibunching) or the two-photon transition (leading to bunching). 
As $\Omega_R$ increases, the linewidths broadens due to power broadening. The spectral lines eventually overlap, leading to $g^{(2)}(0)\simeq 1$. The broadening  scales quadratically (with $\Omega_R^2$) for the two-photon transition and linearly (with $\Omega_R$) for the single-photon transition, which explains the different threshold values of  $\Omega_{R,lim}$ reported in Table~\ref{tab:g2_criteria}.
A similar reasoning applies to dephasing: an increase in $\gamma^*$ also broadens the transitions, yielding analogous threshold values $\gamma^*_{lim}$.
A key point for experiments is that these analytical limits depend on the coupling strength $\Omega_{12}/\Gamma_0\simeq 1/r^3$  where $r$ is the inter-emitter separation. 
In practice, this ratio can be quite large ($\Omega_{12}/\Gamma_0\sim 10-100$), offering substantial flexibility in the accessible range of dephasing rates. 
Eventually, since spectral broadening is weaker for single-photon than for two-photon transitions, the corresponding thresholds are less restrictive for the  $|G\rangle \rightarrow |S\rangle$ than for the $|G\rangle \rightarrow |E\rangle$ transition.
\begin{figure*}
\includegraphics[width=\textwidth]{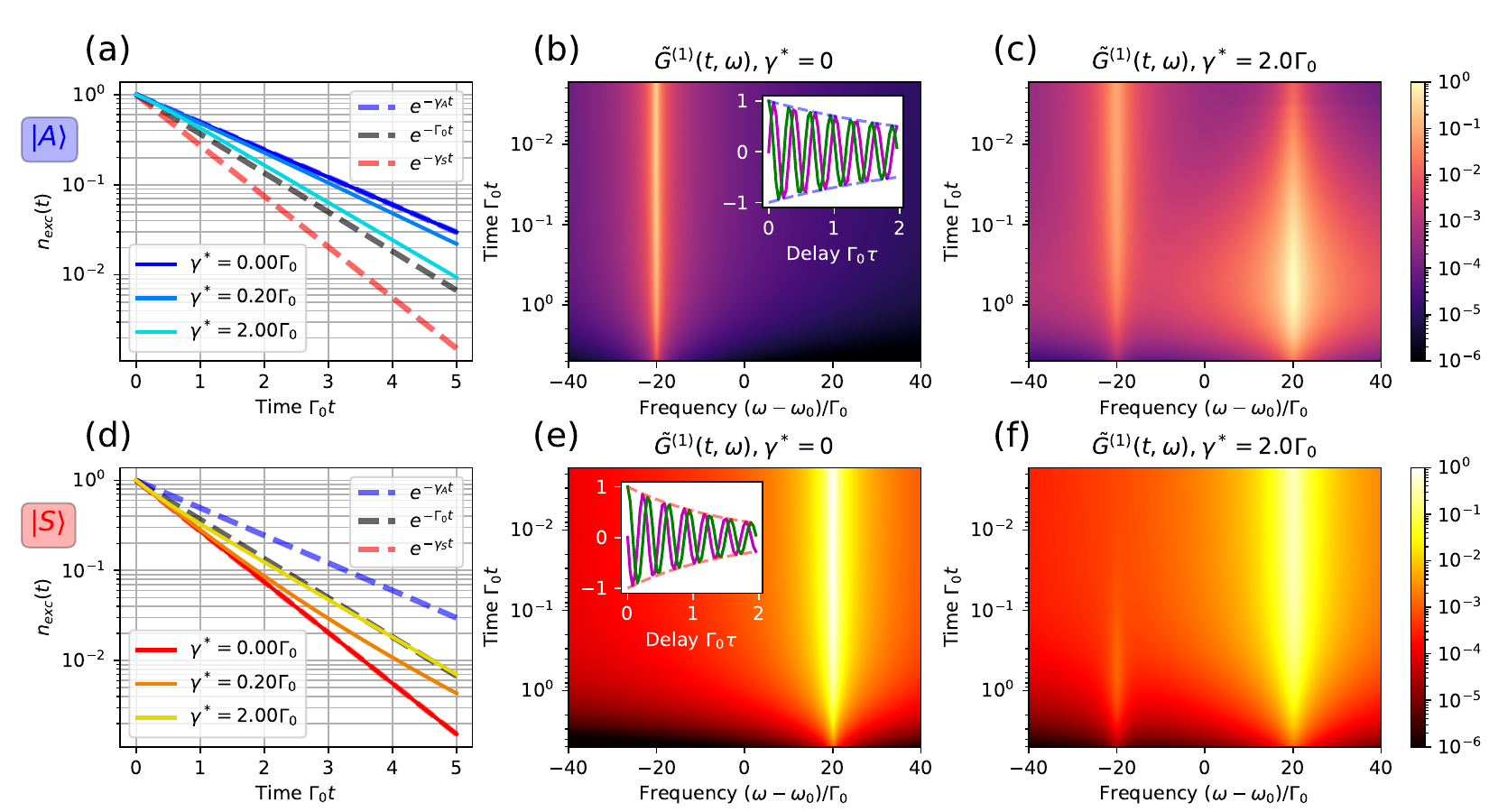}
\caption{\label{fig:g1/main} Study of the dynamics of the system initialized in the first manifold $n_{exc} = 1$.  In (a) [respectively (d)] , we plot $n_{exc}(t)$ starting from the antisymmetric state $|A\rangle$ [respectively $|S\rangle$] for different dephasing rates. The decay is mono-exponential with $\gamma^* = 0$ and bi-exponential otherwise. In (b,c,e,d) we plot the Fourier transform of $G^{(1)}(t, t+\tau)$ with respect to the delay $\tau$ as a function of the frequency and the time $t$. We start from $|A\rangle$, with $\gamma^* = 0$  in (b) and with $\gamma^*=2.0\Gamma_0$ in (c). We start from $|S\rangle$, with $\gamma^* = 0$  in (e) and with $\gamma^*=2.0\Gamma_0$ in (f). The insets show the real \& imaginary parts of $G^{(1)}(t, t+\tau)$ for $\gamma^* = 0$. The emitters are  identical ($\Delta = 0$), placed $r \approx 0.035\lambda$ apart, in an 'H' configuration, with a Debye-Waller/Franck-Condon factor $\alpha = 0.3$, yielding a coupling strength $\Omega_{12} = 20\Gamma_0$ and a dissipative coupling rate $\gamma_{12} = 0.3\Gamma_0$. }
\end{figure*}

\section{\label{sec_free}Free evolution of the coupled emitters}
In this section, we analyze the effect of pure dephasing on the free temporal evolution ($\Omega_R = 0$) of the two coupled emitters.
Indeed, collective effects such as the superradiant burst~\cite{dicke1954coherence,gross1982superradiance} can be revealed by preparing the system in a specific initial state at $t=0$ and monitoring its subsequent evolution. In this section we study two observables: the number of TLS excitations as a function of time and the time-resolved correlations functions of the radiated field $\mathbf{E}_{\text{ZPL}}$.
We study numerically and analytically the free-dynamics using the Liouvillian form of the master equation given in Eq \ref{eq:Liouville}. 
Specifically, we compute the expectation value of any time-dependent observable using the spectral decomposition of the Liouvillian $\mathcal{L}$ and its hermitian conjugate \cite{stefanini2025lindbladme}.
\begin{eqnarray}
    \mathcal{L}[\rho_{R,\mu}] &= \lambda_\mu \rho_{R,\mu} \nonumber \\ 
    \mathcal{L}^\dagger[\rho_{L,\mu}] &= \lambda^*_\mu \rho_{L,\mu},
    \label{eq:SpecLiouv}
\end{eqnarray}
where $\lambda_\mu$ are the eigenvalues, $\rho_{R,\mu}$ and $\rho_{L,\mu}$ are the right and left eigenoperators, normalized such that $\text{Tr}(\rho_{L,\mu}^\dagger\rho_{R,\mu'}) = \delta_{\mu,\mu'}$. 
For a pair of coupled two-level systems, $\mathcal{L}$ has 16 eigenvalues $\lambda_\mu$ and corresponding eigenoperators.
The spectral decomposition of $\mathcal{L}$ allows us to express the density matrix as:
\begin{equation}
    \rho(t) = \sum_\mu e^{\lambda_\mu t}\text{Tr}\left(\rho^\dagger_{L,\mu}\rho_0\right)\rho_{R,\mu},
    \label{eq:rhoLiouv}
\end{equation}
where $\rho_0 = \rho(t=0)$  is the initial density matrix. 
From Eq.~\eqref{eq:rhoLiouv}, the expectation value of any operator $\langle\mathcal{O}(t)\rangle$ can be directly computed, as well as multi-time correlation functions: e.g. $\langle\mathcal{A}(t)\mathcal{B}(t+\tau)\rangle$ using the quantum regression theorem \cite{breuer2002theory}.
This section is organized as follows. In the first part, we prepare the system in a state with only one excitation, leaving the doubly excited state empty. In the second part, we prepare the system in the doubly-excited state $|\psi(t=0)\rangle = |ee\rangle$. In both scenarios, we discuss how pure dephasing modifies the dynamics of the excitation number as well as the first and second-order correlation functions of the emitted field $\mathbf{E}_{\text{ZPL}}$.
\subsection{Single excitation subspace}
In this subsection, we initialize the system in either $|S\rangle$ or $|A\rangle$ and study the dynamics of both the excited state populations $n_{exc}(t)$, and the first-order correlations of the radiated electric field $G^{(1)}(t, t+\tau) = \langle D^\dagger(t)D(t+\tau)\rangle$. In this subsection $D$ defined in Eq. \ref{eq:Dvector} with $\mathbf{\hat{R}}$ a unit vector pointing along x in order to observe the emission of both the superradiant and the subradiant states.
$G^{(1)}(t, t+\tau)$ contains both temporal and spectral information and can be accessed experimentally using an interferometric setup and a detector that records the time of arrival of photons \cite{Belgacem:25}. 
The second-order correlation function $g^{(2)}$ is exactly 0 with our choice of initial states since $|S\rangle$ and $|A\rangle$ can only decay by emitting a single photon, never 2 photons. 

Figures~\ref{fig:g1/main}(a,d) display the time evolution of $n_{exc}(t)$ for initial states $|\psi(0)\rangle = |A\rangle$ and $|\psi(0)\rangle = |S\rangle$ under varying $\gamma^*$.
In all cases, $n_{exc}(0)=1$ and decays to 0. For $\gamma^*=0$, the decay is mono-exponential with rates corresponding to the standard superradiant and subradiant modes: $\Gamma_0 \pm \gamma_{12}$ (indicated by red and blue dashed lines). 
When $\gamma^*>0$, the  decay becomes bi-exponential with rates $\gamma_\pm = \Gamma_0 + \frac{1}{2}(\gamma^*\pm\gamma_{12}^*)$, with $\gamma_{12}^* = \sqrt{{\gamma^*}^2 + 4\gamma_{12}^2}$.
In the limit $\gamma^* \gg 2\gamma_{12}$, the dynamics converge to the independent-emitter decay at rate $\Gamma_0$ (indicated in black dashed lines) irrespective of the initial state.
These features follow naturally from the spectral decomposition of $\mathcal{L}$: the evolution in this subspace is governed by only two eigenmodes associated with $\gamma_\pm$.
For small $\gamma^*$, these eigenmodes correspond mainly to population decay, while at large $\gamma^*$ one of them becomes predominantly coherence-like (see Appendix~\ref{app:spectral_decomposition}).
The crossover occurs near $\gamma^* \simeq \gamma_{12}$.  Since $\gamma_{12}\ll\Omega_{12}$ for closely spaced emitters, this condition can be significantly more stringent than those derived for steady-state observables. 
Let us now  study of the correlation function $G^{(1)}(t, t+\tau)$, which  can be expressed using the general expression for two-time correlations derived in Appendix~\ref{app:spectral_decomposition}:
\begin{align}
    \label{eq:G1_t_tau}
    G^{(1)}(t, t+\tau) &= \sum_{\mu,\mu'} e^{\lambda_\mu t}e^{\lambda_{\mu'}\tau}\biggr[\text{Tr}\left(\rho^\dagger_{L,\mu}\rho_0\right)\nonumber\\
    &\times\text{Tr}\left(\rho^\dagger_{L,\mu'}\rho_{R,\mu}D^\dagger\right) \times\text{Tr}\left(D\rho_{R,\mu'}\right)\biggl].
\end{align}
$G^{(1)}(t, t+\tau)$ is composed of exponential terms in both $t$ and $\tau$, possibly modulated by oscillations if the eigenvalues $\lambda_\mu$ possess  imaginary parts. Although Eq.~\eqref{eq:G1_t_tau} formally contains $16\times16$ terms, only a small subset contribute, depending on the initial-state overlap $\text{Tr}\left(\rho^\dagger_{L,\mu}\rho_0\right)$, and the radiative pathways accessible through $\text{Tr}\left(\rho^\dagger_{L,\mu'}\rho_{R,\mu}D^\dagger\right)\text{Tr}\left(D\rho_{R,\mu'}\right)$ with $D$ defined in Eq. \ref{eq:Dvector} with $\mathbf{\hat{R}}$ a unit vector pointing along x.
The insets in Fig. \ref{fig:g1/main} (b) and (e) show the real (green line) and imaginary (purple line) parts of $G^{(1)}(0, \tau)$ for $|\psi(0)\rangle = |A\rangle$ (b) and $|\psi(0)\rangle = |S\rangle$ (e). The dashed lines show exponential decays at rate $(\Gamma_0 - \gamma_{12})/2$ (blue) and $(\Gamma_0 + \gamma_{12})/2$ (red).
To make the physical content of $G^{(1)}(t, t+\tau)$ more transparent, highlighting its characteristic frequencies and decay rates, we perform a Fourier Transform with respect to the delay $\tau$:
\begin{align}
    \tilde{G}^{(1)}(t,\omega) &= \mathfrak{Re}\int_0^\infty d\tau e^{i\omega\tau}G^{(1)}(t,t+\tau)\nonumber\\
    &= \mathfrak{Re}\sum_{\mu,\mu'} e^{\lambda_\mu t}C_{\mu,\mu'}\int_0^\infty d\tau e^{(i\omega + \lambda_{\mu'})\tau}\nonumber\\
    &= -\mathfrak{Re}\sum_{\mu,\mu'} \frac{e^{\lambda_\mu t}}{i\omega + \lambda_{\mu'}}C_{\mu,\mu'},
    \label{eq:G1spectral}
\end{align}
with $C_{\mu,\mu'} = \text{Tr}\left(\rho^\dagger_{L,\mu}\rho_0\right)\text{Tr}\left(\rho^\dagger_{L,\mu'}\rho_{R,\mu}D^\dagger\right)\text{Tr}\left(D\rho_{R,\mu'}\right)$. From  Eq. (\ref{eq:G1spectral}), it is clear that $\tilde{G}^{(1)}(t,\omega)$ is peaked at frequencies corresponding to the imaginary part of  $\lambda_\mu$ [responsible for the oscillations in $G^{(1)}(t, t+\tau)$], and the width of these peaks corresponds to the decay rate of these oscillations (associated to the real parts of $\lambda_\mu$).\\
The middle column of Fig. \ref{fig:g1/main} shows $\tilde{G}^{(1)}(t,\omega)$ without dephasing, for $|\psi(0)\rangle=|A\rangle$ (b) and $|S\rangle$ (e). When the system is initially prepared in $|A\rangle$ (respectively $|S\rangle$) $\tilde{G}^{(1)}(t,\omega)$ is a Lorentzian centered on $-\Omega_{12}$ (respectively $+\Omega_{12}$) with a full width at half-maximum (FWHM) $\Gamma_0 - \gamma_{12}$ (respectively $\Gamma_0 + \gamma_{12}$) whose amplitude decays exponentially in time at rate $\Gamma_0 - \gamma_{12}$ (respectively $\Gamma_0 + \gamma_{12}$). Indeed, when $\gamma^*=0$, both $|A\rangle\langle A\vert $ and $|S\rangle\langle S\vert $ project on a single Liouvillian eigenstate. Hence, the emission  frequency remains constant at $\pm\Omega_{12}$. 
In contrast, for finite dephasing $\gamma^*\neq 0$, neither $|A\rangle\langle A\vert $ nor $|S\rangle\langle S\vert $ project on a single eigenstate of the full Liouvillian anymore.
Consequently, $\tilde{G}^{(1)}(t,\omega)$ displays peaks at both $\omega=\pm \Omega_{12}$ regardless of the initial state: see Fig. \ref{fig:g1/main}  (c) and  (f).
This behavior reflects the mixing of symmetric and antisymmetric populations induced by dephasing. When the system starts in $\vert A \rangle$, population gradually transfers to $\vert S \rangle$ producing an emerging peak at $\omega=+\Omega_{12}$ which eventually dominates due to the stronger radiative character of $\vert S \rangle$.
The reverse process occurs for initial $\vert S \rangle$ though the antisymmetric contribution remains weak.
This population transfer becomes significant when $\gamma^*\geq \gamma_{12}$, setting a stringent condition for the observation of collective effects.

\subsection{Two-excitations subspace}
\begin{figure*}
\includegraphics[width=\textwidth]{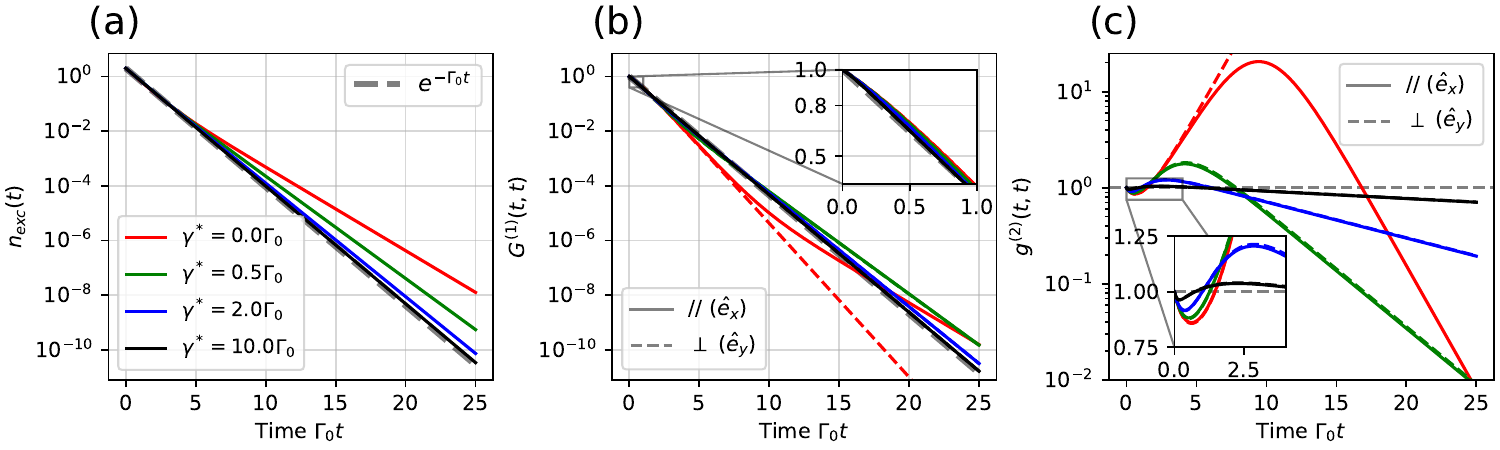}
\caption{\label{fig:pulsed_g2/main} Dynamics of the system, starting from the doubly-exited state $|E\rangle$. In (a) we plot the excited state population $n_{exc}$ as a function of time, for different dephasing rates $\gamma^*$.  In (b) we plot the un-normalized first-order correlation function at zero delay $G^{(1)}(t,t)$. The inset shows a zoom on the first instants of the evolution, from $\Gamma_0t=0$ to $0.5$.   In (c) we plot the normalized second-order correlation function at zero-delay $g^{(2)}(t, t)$. The inset shows a zoom on the first instants of the evolution. In all cases, the different colors (but grey) corresponds to different values of $\gamma^*$.  In (b) and (c) we consider two different detection schemes: along the inter-emitter axis ($//$, plain lines) and along a perpendicular bisector of the inter-emitter axis ($\perp$, dashed lines). For comparison, in all panels we represent the dynamics of independent emitters in dashed grey line. The emitters are identical ($\Delta = 0$), placed $r \approx 0.035\lambda$ apart, in an 'H' configuration, with a Debye-Waller/Franck-Condon factor $\alpha = 0.3$, yielding a coupling strength $\Omega_{12} = 20\Gamma_0$ and a dissipative coupling rate $\gamma_{12} = 0.3\Gamma_0$.}
\end{figure*}
We now consider the case where the system is initially prepared in the fully inverted, doubly excited state $\rho_0=|E\rangle\langle E|$.
Unlike the partially excited case discussed earlier, this initial state is separable, with both emitters excited independently.
Since two photons must be emitted for the system to relax to the ground state, it is natural to investigate not only the first-order field correlations $G^{(1)}$, the excited-state population $n_{exc}$, but also the normalized second-order correlation function $g^{(2)}$ defined as:
\begin{equation}
\label{eq:g2_t_tau}
    g^{(2)}(t, t+\tau) = \frac{
        \langle D^\dagger(t)D^\dagger(t+\tau)D(t+\tau)D(t)\rangle
    }{
        \langle D^\dagger(t)D(t)\rangle\langle D^\dagger(t+\tau)D(t+\tau)\rangle
    }.
\end{equation}
In Fig. \ref{fig:pulsed_g2/main} (a), we plot the decay of excitation number as a function of time for several dephasing rates.
At short times $t \lesssim 1/\Gamma_0$, the decay follows that of uncoupled emitters (gray dashed line)  since the doubly excited manifold contains only a single state and is therefore insensitive to dephasing.
At longer times, the decay slows down and proceeds at the smallest population decay rate $\Gamma_0 + \frac{1}{2}(\gamma^* - \gamma_{12}^*)$ (the smallest population decay rate in our system).
For large $\gamma^*$, so this rate approaches $\Gamma_0$, leading to a monoexponential decay similar to that of independent emitters.
Panel (b) shows the first-order field correlation $G^{(1)}(t,t)$, proportional to the detected intensity.
Solid lines correspond to detection along the inter-emitter axis (along $\mathbf{\hat{e}_x}$), while dashed lines correspond to perpendicular detection (along $\mathbf{\hat{e}_y}$).
At early times ($t \lesssim 1/\Gamma_0$), $G^{(1)}(t,t)$ slightly exceeds  the monoexponential decay at rate $\Gamma_0$ (dashed gray line) expected for uncoupled emitters.
This small enhancement corresponds to the onset of a superradiant burst \cite{gross1982superradiance}, which would become more pronounced in systems with more emitters.
At later times, the decay of $G^{(1)}(t,t)$ follows the slowest population decay rate $\Gamma_0 + \frac{1}{2}(\gamma^* - \gamma_{12}^*)$, except for perpendicular detection when $\gamma^* = 0$ (dashed red line).
In this case, the antisymmetric state does not radiate in the perpendicular direction, and only the superradiant decay at rate $\Gamma_0 + \gamma_{12}$ is observed.
As $\gamma^*$ increases, the decay approaches a single exponential at rate $\Gamma_0$, and the dependence on detection direction disappears.
Thus, at large $\gamma^*$, the system behaves as a pair of independent emitters.
Panel (c) presents the normalized second-order correlation function at zero delay, $g^{(2)}(t,t)$, for detection along (solid lines) and perpendicular to (dashed lines) the inter-emitter axis.
The temporal evolution of $g^{(2)}(t,t)$ typically exhibits a dip below $1$, followed by a rise above $1$, a maximum, and finally a decay toward $0$.
The only exception occurs for perpendicular detection with $\gamma^* = 0$ (dashed red line), where $g^{(2)}(t,t)$ diverges instead of showing a maximum.
As $\gamma^*$ increases, both the depth of the initial dip and the height of the subsequent maximum decrease, and these features occur at earlier times.
Moreover, the difference between parallel and perpendicular detection gradually vanishes with increasing $\gamma^*$, and for sufficiently strong dephasing, $g^{(2)}(t,t)$ tends to $1$ at all times.
To understand these dynamics, $g^{(2)}(t,t)$ can be expressed in terms of the density matrix populations:
\begin{align}
    g^{(2)}(t,t) &= \frac{\rho_{ee,ee}}{(\rho_{ee,ee} + \rho_{S,S}\cos^2{(\phi/2)} + \rho_{A,A}\sin^2(\phi/2))^2} \nonumber \\
    \label{eq:g2_t_t_pop_approx}
     &\approx \frac{\rho_{ee,ee}}{(\rho_{ee,ee} + \rho_{S,S} + \frac{\phi^2}{4}(\rho_{A,A}-\rho_{S,S}))^2},
\end{align}
with $\rho_{S,S} = \langle S | \rho(t) | S \rangle$, $\rho_{A,A} = \langle A | \rho(t) | A \rangle$.  The phase difference between the fields radiated by each emitter in the direction of the detector is noted $\phi = k\mathbf{\hat{R}}\cdot\mathbf{r}$.
$\phi = 0$ for perpendicular detection and $\phi=kr$ for parallel detection, justifying the small-angle approximation $\phi \ll 1$ done in Eq. (\ref{eq:g2_t_t_pop_approx}).
Since the doubly excited population simply evolves as $\rho_{ee,ee} = e^{-2\Gamma_0 t}$, the behavior of $g^{(2)}(t,t)$ is determined by the denominator, which scales as $G^{(1)}(t,t)^2$.
When $G^{(1)}(t,t)^2 < e^{-2\Gamma_0t}$, $g^{(2)}(t,t)>1$ and conversely if $G^{(1)}(t,t)^2 > e^{-2\Gamma_0t}$ then $g^{(2)}(t,t) < 1$.
Thus, the short-time bump in $G^{(1)}(t,t)$ directly corresponds to the initial dip in $g^{(2)}(t,t)$.
Physically, photon emission is enhanced compared to independent emitters, but photons are less likely to arrive in pairs.
At intermediate times, emission is dominated by the symmetric state, which decays faster than $\Gamma_0$, causing the denominator of Eq.~(\ref{eq:g2_t_tau}) to decrease faster than the numerator.
This results in $g^{(2)}(t,t) > 1$, indicating photon bunching.
In this regime, few photons are emitted, but $\rho_{S,S}^2 < \rho_{ee,ee}$.
For $\gamma^*=0$ and perpendicular detection, $G^{(1)}(t,t)$ decays at the superradiant rate $\Gamma_0+\gamma_{12}$ at late times so $g^{(2)}(t,t)$ diverges exponentially at rate $2\gamma_{12}$.
In all other cases, at long enough times, $G^{(1)}(t,t)$ decays exponentially at a rate given by the smallest population decay rate $\Gamma_0 + \frac{1}{2}(\gamma^* - \gamma_{12}^*)$, so $g^{(2)}(t,t)$ eventually decays to zero exponentially at rate $\gamma_{12}^*-\gamma^*$. This corresponds to antibunched emission.
When $\gamma^* \rightarrow \infty$, the above quantity $\gamma_{12}^*-\gamma^*\rightarrow0$ so $g^{(2)}(t,t)$ stays equal to $1$ at all times which is the expected result for a pair of uncoupled emitters.
Here again the limiting dephasing rate to observe a dip and a maximum in $g^{(2)}(t,t)$  is on the order of the dissipative coupling rate $\gamma_{12}$.
Finally, Eq. (\ref{eq:g2_t_t_pop_approx}) clarifies how dephasing suppresses directional dependence.
The detection geometry enters only through $\phi$, while $\gamma^*$ couples the symmetric and antisymmetric states, equalizing their populations.
Consequently, for large  $\gamma^*$ the term $\frac{\phi^2}{4}(\rho_{A,A}-\rho_{S,S})$  rapidly vanishes for any $\phi$, erasing the distinction between detection directions

\section{Conclusion}

In summary we have conducted a systematic study of the competition between pure dephasing and collective effects for a  system  composed of $N=2$ emitters. We modeled the dephasing using an independent dephasing rate $\gamma^*$ acting separately on each emitter. 
Such dephasing can originate from various mechanisms, including charge fluctuations~\cite{itakura2003dephasing}, interactions with phonons~\cite{krummheuer2002theory,wiercinski2023phonon}, or collisions in atomic vapors~\cite{maki1989influence}.
Our analysis has primarily relied on the local master equation, for which the temperature dependence only enters through the dephasing rate $\gamma^*(T)$. We have also discussed the corresponding global description and shown that, in the experimentally relevant regime $\hbar\Omega_{12}\ll k_B T$, the difference between the two approaches is negligible.

In all cases, strong dephasing suppresses the entanglement of the two-body density matrix $\rho$, rendering it factorizable as $\rho = \rho_1 \otimes \rho_2$, and leading to independent emission.
However, the magnitude of $\gamma^*$ required to achieve this depends both on the system’s excitation scheme and on the observable under consideration.
We identified and summarized in Fig.~\ref{fig:schemaexp}(b) the critical dephasing rates beyond which collective signatures vanish in four different observables.
For all studied cases, observables in the steady-state regime were associated with limiting $\gamma^*$ values that scale as powers of the coherent coupling $\Omega_{12}$, which can be very large at short inter-emitter separations.
 In contrast, freely evolving observables exhibited limiting $\gamma^*$ values on the order of the dissipative coupling rate $\gamma_{12}$, itself comparable to $\Gamma_0$. 
Beyond setting the overall dephasing strength through $\gamma^*(T)$, temperature also determines the validity of the local description through the ratio $\hbar\Omega_{12}/k_B T$. For experimentally relevant parameters, this ratio is typically small, supporting the use of the local master equation adopted throughout our analysis. However, ongoing efforts to engineer strongly coupled and low-temperature quantum emitters are progressively approaching regimes where $\hbar\Omega_{12}/k_B T$ is no longer negligible, making the global description increasingly relevant.
 As a consequence, these results provide useful guidelines for assessing the robustness of collective effects against pure dephasing and for designing experiments and protocols that aim to exploit collective emission phenomena in applications such as on-demand quantum memories~\cite{ballantine2021quantum,rubies2022photon,fayard2023optical}.

In future work, we plan to extend this analysis to larger ensembles of emitters ($N>2$), using the global master equation framework to investigate regimes where $\hbar\Omega_{12}/k_B T$ is no longer negligible. This will become particularly relevant for strongly coupled and low-temperature emitters, where the interplay between collective dynamics and temperature-dependent phonon-induced transitions can no longer be captured by the local description. Extending this framework to larger ensembles of emitters would then open the way to the study of genuine many-body phenomena, such as many-body localization and highly excited superradiance~\cite{fayard2021many,masson2020many,ferioli2023non,ferioli2025emergence}.
The main challenge in this direction lies in the exponential growth of the Hilbert space, which scales as $2^N$, making both analytical and numerical treatments increasingly demanding. Several approaches can mitigate this computational complexity. First, simplified models such as the Dicke model~\cite{shammah2018open}, where permutation symmetry reduces the problem size to $\mathcal{O}(N^3)$, can provide valuable insight. Second, truncating the Hilbert space to the lowest or highest excitation manifolds allows one to capture early-time dynamics~\cite{robicheaux2021theoretical,masson2022universality} or few-excitation regimes. Third, cumulant expansion techniques can approximate the dynamics by truncating higher-order operator correlations~\cite{ferioli2021laser,rubies2023characterizing}. Finally, throughout this work we have considered Markovian photonic and phononic environments. An exciting further direction would therefore be to explore the competition between collective effects and pure dephasing in engineered non-Markovian baths, such as structured phononic~\cite{krummheuer2002theory,wiercinski2023phonon,hall2025controlling} or photonic~\cite{lambropoulos2000fundamental,Fayard22,bouscal2024systematic} reservoirs.
\section*{Acknowledgements}
We acknowledge C. Diederich, Y. Chassagneux, C. Voisin, C. Mayer, E. Cassette, G. Trippe-Allard, E. Deleporte for important discussions.
N.F acknowledges funding from the Agence National de la Recherche (ANR, project CSUPER2).

\appendix

\section{\label{app:steady_state_lorentzians}Steady-state lorentzians}
Over a broad range of parameters, the excitation spectrum $n_{exc}(\omega)$ described in section~\ref{sec_EP} can be accurately described by three Lorentzian peaks, denoted $\mathcal{L}_A$, $\mathcal{L}_E$, and $\mathcal{L}_S$ (see panel (d) in Fig.~\ref{fig:ple/main}). These correspond respectively to the transitions $|G\rangle \rightarrow |A\rangle$, $|G\rangle \rightarrow |E\rangle$, and $|G\rangle \rightarrow |S\rangle$. In this appendix, we first motivate the use of $n_{exc}$ to model the red-shifted emission intensity, then we focus on the theoretical description of $\mathcal{L}_E$ and $\mathcal{L}_S$, which are both accessible in the zero-detuning regime considered in our analytical treatment.
\subsection{\label{app:redshifted_intensity}Red-shifted emission intensity}
 We can split the Lindblad term associated to the optical decay (second right-hand side term)   of the master equation given in Eq.\ref{eq:Mastereq_dephasing_local} into two contributions, one from the ZPL and one from the red-shifted emission:
\begin{align}
    \mathcal{J}_{tot}[\rho] = \mathcal{J}_{\text{ZPL}} [\rho]+ \mathcal{J}_{vib}[\rho]
\end{align}
with:
\begin{align}
    \mathcal{J}_{tot}[\rho] &= \sum_{i,j = 1}^{2}\frac{\gamma_{ij}}{2}L_{\sigma_i, \sigma_j}[\rho]\\
    \mathcal{J}_{\text{ZPL}}[\rho] &= \frac{1}{2}\alpha\Gamma_0\sum_{i,j=1}^2 L_{\sigma_i, \sigma_j}[\rho]\\
    \mathcal{J}_{vib}[\rho] &= \frac{1}{2}(1-\alpha)\Gamma_0\sum_{i=1}^2 L_{\sigma_i}[\rho].
\end{align}
This splitting permits to compute the rate of red-shifted photon emission trough $\langle G| \mathcal{J}_{vib}[\rho] | G \rangle - \langle E| \mathcal{J}_{vib}[\rho] | E \rangle $. $\langle G| \mathcal{J}_{vib}[\rho] | G \rangle$ is the rate at which the ground state population increases due to red-shifted emission, while  $- \langle E| \mathcal{J}_{vib}[\rho] | E \rangle$ is the rate at which the doubly-excited state population decreases due to red-shifted emission. Since there is no emission from the doubly-excited state directly to the ground state ($\langle G| \mathcal{J}_{vib}[|E\rangle\langle E|] | G \rangle = 0$), the first term contains all red-shifted emission from the 1-excitation manifold to the ground state, while the second contains all the red-shifted emission from the doubly-excited state to the 1-excitation manifold. Performing the computation yields $\langle G| \mathcal{J}_{vib}[\rho] | G \rangle - \langle E| \mathcal{J}_{vib}[\rho] | E \rangle = (1-\alpha)\Gamma_0(2\rho_{ee,ee} + \rho_{eg,eg} + \rho_{ge,ge}) \propto n_{exc}$. Thus the total rate of red-shifted photon emission is proportional to $n_{exc} = (2\rho_{ee,ee} + \rho_{eg,eg} + \rho_{ge,ge})$. Finally, since the vibrational energy levels involved in the transitions are extremely short lived, the red-shifted emission between each emitter is incoherent and can't interfere. Then, aside from the usual dipole radiation pattern, this emission is not directional and the detected intensity is proportional to $n_{exc}$ regardless of the detection scheme.
\subsection{Superradiant peak}
For the three spectral peaks to be well resolved, the splitting $\Omega_{12}$ must exceed the linewidths, which depend in particular on $\Omega_R$ and $\gamma^*$ (power broadening and homogeneous broadening, respectively).
To derive the analytical expressions of these Lorentzians, we expand our theoretical spectrum found using SymPy for $\Delta = 0$ and available in~\cite{lumin_analytics}, around $\omega = 0$ (corresponding to the $|G\rangle \rightarrow |E\rangle$ two-photon transition) and $\omega = \Omega_{12}$ (corresponding to the $|G\rangle \rightarrow |S\rangle$ transition). By matching these series expansions with that of a Lorentzian function, we obtain the width and amplitude of each peak.
Further simplification can be achieved by assuming $\Omega_{12} \gg \Gamma_0, \gamma_{12}, \Omega_R, \gamma^*$, a regime in which the three peaks are well separated. Under this condition, the superradiant peak $\mathcal{L}_S(\omega)$ takes the form:
\begin{equation}
    \mathcal{L}_S(\omega) \approx \frac{2\Omega_R^2\left(\Gamma_0 + \gamma^* - \gamma_{12}\right)\left(\Gamma_0 + \gamma^* + \gamma_{12}\right)}{A_S+B_S+C_S}
\end{equation}
where we introduced 
\begin{align}
   &A_S=4\Omega_R^2 (\Gamma_0 + \frac{3}{4} \gamma^* - \gamma_{12})\left(\Gamma_0 + \gamma^* + \gamma_{12}\right) \\ 
    &B_S=4\left(\omega-\Omega_{12}\right)^2\left(\Gamma_0^2 + \Gamma_0 \gamma^* - \gamma_{12}^2\right)\\
    &C_S= \left( \Gamma_0 + \gamma^* + \gamma_{12}\right)^2\left(\Gamma_0^2 + \Gamma_0\gamma^* - \gamma_{12}^2\right)
\end{align}
to simplify the writing.
In the limit of weak driving field ($\Omega_R \ll \Gamma_0$), this expression reduces to:
\begin{align}
    \label{eq:steady_state_superradiant_lorentzian}
    \mathcal{L}_S(\omega) \approx
    \frac{2\Omega_R^2\left(\Gamma_0 + \gamma^* - \gamma_{12}\right)\left(\Gamma_0 + \gamma^* + \gamma_{12}\right)}{B_S+C_S}
\end{align}
This Lorentzian has a full width given by $\Gamma^2=(\Gamma_0+\gamma^*+\gamma_{12})^2$, and  reaches its maximum at $\omega=\Omega_{12}$ equal to:
\begin{align}
    \mathcal{L}_S(\Omega_{12})=\frac{2\Omega_R^2(\Gamma_0+\gamma^*-\gamma_{12})}{(\Gamma_0+\gamma^*+\gamma_{12})(\Gamma_0^2+\Gamma_0\gamma^*-\gamma_{12}^2)}.
\end{align}
In the absence of pure dephasing ($\gamma^* = 0$), we recover the well-known results
$\Gamma=\Gamma_0+\gamma_{12}$, with an amplitude $\mathcal{L}_S(\Omega_{12})=\frac{2\Omega_R^2}{(\Gamma_0+\gamma_{12})^2}$.
\subsection{Two-photon excitation peak}
We now turn to the analytical description of the doubly excited peak.
We first consider the situation where pure dephasing is negligible, i.e., $\gamma^* \ll \Omega_R, \Omega_{12}$. In addition, we assume that $\Omega_{12} \gg \Gamma_0$ (so that the superradiant and doubly excited peaks are spectrally resolved).
In this regime, the excitation spectrum around $\omega = 0$ (corresponding to the $|G\rangle \rightarrow |E\rangle$ two-photon transition) can be approximated by a Lorentzian with an amplitude at $\omega = 0$ given by:
\begin{equation}
\mathcal{L}_E(\omega=0)\simeq \frac{2(X^2+2X^4)}{4(X^2+X^4+Y^2)}
\end{equation}
where we have introduced the normalized parameters $X = \Omega_R / \Gamma_0$ and $Y = \Omega_{12} / \Gamma_0$.
Three regimes can be distinguished:
\begin{itemize}
\item \textbf{$X \gg Y$:} the Rabi frequency dominates, and the doubly excited peak becomes saturated, with $\mathcal{L}_E(\omega = 0) \sim 1$.
\item \textbf{$Y \gg X \gg 1$:} in this regime, the population of the doubly excited state dominates the signal at $\omega = 0$. The amplitude scales as $\mathcal{L}_E(\omega = 0) \sim X^4 / Y^2$, i.e., it grows proportionally to the \emph{square} of the incident field intensity.
\item \textbf{$Y \gg 1 \gg X$:} here, the contribution at $\omega = 0$ originates mainly from the tail of the superradiant peak. The amplitude scales as $\mathcal{L}_E(\omega = 0) \sim X^2 / 2Y^2$, i.e., it increases \emph{linearly} with the incident field intensity.
\end{itemize}

\subsection{\label{app:steady_state_saturation}Linear and quadratic scalings of $n_{exc}(\omega=0)$ as a function of the field intensity. }
Let us now detail the derivation of the limiting parameter $\gamma^*_{lim}$that marks the transition between the linear and superlinear regimes of the  saturation curves represented in panel (c) in Fig. \ref{fig:ple/main}.
Setting $\omega = 0$ in our analytical expression for $n_{exc}$, we find
\begin{equation}
    n_{exc} = \Omega_R^2\frac{\Omega_R^4A_4 + \Omega_R^2A_2 + A_0}{\Omega_R^6 B_6 + \Omega_R^4 B_4 + \Omega_R^2 B_2 + B_0},
\end{equation}
with $A_0,A_2,A_4,B_0,B_2,B_4,B_6$ different coefficients of the numerator and the denominator of $n_{exc}$ corresponding to different powers of $\Omega_R$.
To analyze whether the saturation curve exhibits a linear or quadratic dependence at low excitation intensity, we expand $n_{\mathrm{exc}}$ as a power series in $\Omega_R^2 / \Gamma_0^2$, retaining terms up to second order:
\begin{equation}
    n_{exc}\approx \Omega_R^4\left(\frac{A_2}{B_0}  - \frac{A_0B_2}{B_0^2}\right) + \Omega_R^2\frac{A_0}{B_0}.
\end{equation}
The coefficient of the linear term $A_0/B_0$ reads:
\begin{equation}
    \frac{A_0}{B_0} = \frac{2(\Gamma_0 + \gamma^* + \gamma_{12})(\Gamma_0 + \gamma^* - \gamma_{12})}{(\Gamma_0^2 + \Gamma_0\gamma^* - \gamma_{12}^2)(4\Omega_{12}^2 + (\Gamma_0 + \gamma^* + \gamma_{12})^2)}
\end{equation}
is plotted as the dashed lines in Fig.~\ref{fig:ple/main} panel (c).
This expression coincides with that of $\mathcal{L}_S(\omega=0)$ given in Eq.~(\ref{eq:steady_state_superradiant_lorentzian}), indicating that the linear contribution originates from the tail of the superradiant peak.
This explains the presence of a linear scaling region even in absence of dephasing.
The coefficient in front of the quadratic term has a much longer expression. In the large coupling and large dephasing limit $\Omega_{12}, \gamma^* \gg \Gamma_0$, it can be simplified as:
\begin{equation}
    \frac{A_2}{B_0} - \frac{A_0B_2}{B_0^2} \approx 4\frac{4\Omega_{12}^2\Gamma_0 - {\gamma^*}^3}{\Gamma_0^2\gamma^*\left(4\Omega_{12}^2 + {\gamma^*}^2\right)^2}.
\end{equation}
This term vanishes for $\gamma^* = (4\Omega_{12}^2\Gamma_0)^\frac{1}{3}$, which defines the limiting value $\gamma^*_{lim}$.

\subsection{The large dephasing limit}
In the regime where $\gamma^*$ exceeds all other characteristic rates, the superradiant peak remains Lorentzian, centered on $\omega=\Omega_{12}$  with an amplitude $\mathcal{L}_S(\Omega_{12})\sim 2\Omega_R^2/(\Gamma_0\gamma^*)$ and a width $\Gamma=\gamma^*/2$. 
In this limit, the broad superradiant Lorentzian completely overlaps the doubly excited peak, rendering it invisible in the spectrum.

\section{\label{app:g2_0_thresholds} $g^{(2)}(0)$ Thresholds}
In this appendix we derive the  threshold values of $\Omega_R$ (for $\gamma^* = 0$) and of $\gamma^*$ (in the limit $\Omega_R \rightarrow 0$) above which $g^{(2)}(0)$ resembles that  of independent emitters as discussed in section~\ref{section_g2}.
We consider two excitation schemes: (i) resonant excitation ($\omega = \omega_0$) and (ii) excitation at the superradiant frequency ($\omega - \omega_0 = \Omega_{12}$).
From Eq.~(\ref{eq:G2_value}) of the main text, we have:
\begin{align}
    g^{(2)}(0)= \frac{\rho_{ee,ee}}{(\rho_{ee} + \mathfrak{Re}(\rho_{ge,eg}))^2},
\end{align}
with $\rho_{ee} = \rho_{ee,ee} + \rho_{eg,eg} = \rho_{ee,ee} + \rho_{ge,ge}$ the excited state population of a single emitter, obtained by taking the partial trace over the degrees of freedom of the other emitter. Both emitters are identical and under the same excitation, thus they have the same excited state population.
Assuming $\rho_{ee} \gg \mathfrak{Re}(\rho_{ge,eg})$ (valid when $\gamma^*$ or $\Omega_R$ are large compared to other quantities) the behavior of $g^{(2)}(0)$ is mainly governed by the ratio $\rho_{ee,ee} / \rho_{ee}^2$.
In particular, for a separable density matrix, $\rho_{ee,ee} = \rho_{ee}^2$ and $g^{(2)}(0)=1$.
Therefore, to determine the thresholds for $\gamma^*$ and $\Omega_R$, we  focus on the ratio $\rho_{ee,ee}/\rho_{ee}^2$.

\subsection{Excitation at $\omega=\omega_0$}
When exciting the two-photon resonance, the population of the doubly excited state becomes significant, leading to $\rho_{ee,ee} > \rho_{ee}^2$. To quantify this effect, we define the thresholds in $\Omega_R$ and $\gamma^*$ as the values for which $\rho_{ee,ee} =2\rho_{ee}^2$, capturing the onset of significant double excitation.
\subsubsection{$\Omega_R$ Threshold}
We first consider the case $\gamma^* = 0$.
The exact steady-state populations are given by:
\begin{align}
    \rho_{ee,ee} &= \frac{\Omega_R^4}{4\Omega_R^4 + \Gamma_0^2((\Gamma_0+\gamma_{12})^2 + 4\Omega_R^2 + 4\Omega_{12}^2)},\\
    \rho_{ee} &= \frac{\Gamma_0^2\Omega_R^2+2\Omega_R^4}{4\Omega_R^4 + \Gamma_0^2((\Gamma_0+\gamma_{12})^2 + 4\Omega_R^2 + 4\Omega_{12}^2)}.
\end{align}
Assuming $\Omega_{12} \gg \Gamma_0$, we can simplify the denominators as follows:
\begin{align}
    \rho_{ee,ee} &= \frac{1}{4}\frac{\Omega_R^4}{\Omega_R^4 + \Gamma_0^2(\Omega_R^2 + \Omega_{12}^2)},\\
    \rho_{ee} &= \frac{1}{4}\frac{\Gamma_0^2\Omega_R^2+2\Omega_R^4}{\Omega_R^4 + \Gamma_0^2(\Omega_R^2 + \Omega_{12}^2)}.
\end{align}
Hence, the threshold condition $\rho_{ee,ee} =2\rho_{ee}^2$
can be cast as a  second order polynomial in $\Omega_R^2$:
\begin{equation}
4\Omega_R^4 + 4\Gamma_0^2\Omega_R^2 + 2\Gamma_0^4 - 4\Gamma_0^2\Omega_{12}^2 = 0.
\end{equation}
The only positive solution is:
\begin{equation}
    \Omega_R^2 = \frac{-\Gamma_0 + 2\Gamma_0|\Omega_{12}|}{2} =\Gamma_0|\Omega_{12}|,
\end{equation}
resulting in the threshold condition $\Omega_R = \sqrt{\Gamma_0|\Omega_{12}|}$.
\subsubsection{$\gamma^*$ Threshold}
The exact analytical expressions for $\rho_{ee}$ and $\rho_{ee,ee}$ are complex and not easily tractable. We therefore use simplified forms, valid in the regime where $\Omega_{12}, \gamma^* \gg \Gamma_0$ and $\Omega_R \ll\Gamma_0$:
\begin{align}
    \rho_{ee,ee} &\approx \Omega_R^4\frac{4\Omega_{12}^2\Gamma_0\gamma^* + {\gamma^*}^4}{(\Gamma_0\gamma^*)^2(4\Omega_{12}^2 + {\gamma^*}^2)^2}\\
    \rho_{ee} &\approx \Omega_R^2\frac{{\gamma^*}^2}{\Gamma_0\gamma^*(4\Omega_{12}^2 + {\gamma^*}^2)}.
\end{align}
We note that $\rho_{ee,ee}$ can be written as $\rho_{ee,ee} = \rho_{ee}^2$ plus an additional term that becomes significant at finite $\gamma^*$.
Applying the threshold condition $\rho_{ee,ee} = 2\rho_{ee}^2$ yields:
\begin{align}
    {\gamma^*}^3 &=4\Gamma_0\Omega_{12}^2.
\end{align}
This defines the dephasing threshold separating the coherent two-photon regime from the incoherent one: $\gamma^*_{lim} =(4\Omega_{12}^2\Gamma_0)^{1/3}$.
\subsection{Excitation at $\omega-\omega_0=\Omega_{12}$}
When exciting the $|G\rangle\rightarrow|S\rangle$ transition the doubly excited state population is weak and verifies $\rho_{ee,ee} < \rho_{ee}^2$. Consequently, the threshold condition differs from the one chosen for the $|G\rangle\rightarrow|E\rangle$ transition.
\subsubsection{$\Omega_R$ Threshold}
In this case, we use a slightly different condition, because the usual approximation $\rho_{ee} \gg \mathfrak{Re}(\rho_{ge,eg})$ does not hold. Intuitively, since the state being excited is a superposition involving both emitters, the coherence $\rho_{ge,eg}$ can be comparable in magnitude to the populations.
Therefore, we use the following threshold  condition: $\rho_{ee,ee} = \frac{1}{2}(\rho_{ee} + \mathfrak{Re}(\rho_{ge,eg}))^2$ (which is equivalent to the condition $ g^{(2)}(0)=\frac{1}{2}$).
 Setting the condition $g^{(2)}(0)= 1/2$ into:
\begin{align}
    g^{(2)}(0)=\frac{\Omega_R^2(\Omega_R^2+4\Omega_{12}^2) + (\Gamma_0+\gamma_{12})^2\Omega_{12}^2}{(\Omega_R^2 + 4\Omega_{12}^2)^2}.
\end{align}
leads to the threshold $\Omega_R=2|\Omega_{12}|$.

\subsubsection{$\gamma^*$ Threshold}
A strong dephasing rate $\gamma^*$ rapidly destroys quantum coherences. We  therefore define the following threshold condition $\rho_{ee,ee} = \frac{1}{2}\rho_{ee}^2$.
We employ  approximations valid in the regime $\Omega_{12},\gamma^*\gg\Gamma_0$ and $\Omega_R \ll \Gamma_0$:
\begin{align}
    \rho_{ee,ee} &\approx \Omega_R^4\frac{1}{\Gamma_0^2(16\Omega_{12}^2 + {\gamma^*}^2)}\\
    \rho_{ee} &\approx \Omega_R^2\frac{1}{\Gamma_0\gamma^*}.
\end{align}
From these expressions, we obtain the threshold condition $\gamma^*_{lim} \approx 4|\Omega_{12}|$.
\section{\label{app:spectral_decomposition}Spectral decomposition of the liouvillian $\mathcal{L}$}
In this appendix we describe the eigenvalues $\lambda_\mu$ and left and right eigenoperators $\rho_{L,\mu}$ and $\rho_{R,\mu}$ of the Liouvillian relevant to the study of the free dynamics described in section~\ref{sec_free}. The exact analytical expression for the eigenvalues and eigenoperators were found in the $\Delta = 0$ case.
\subsection{Ground state}
We label the eigenstate corresponding to the ground state with $\mu=G$. In the basis $\left\{|gg\rangle,|eg\rangle,|ge\rangle,|ee\rangle\right\}$     
\begin{align}
    \rho_{R,G} = \begin{bmatrix}
        1 & 0 & 0 & 0 \\
        0 & 0 & 0 & 0 \\
        0 & 0 & 0 & 0 \\
        0 & 0 & 0 & 0
    \end{bmatrix}
    , \
    \rho_{L,G}^\dagger = \begin{bmatrix}
        1 & 0 & 0 & 0 \\
        0 & 1 & 0 & 0 \\
        0 & 0 & 1 & 0 \\
        0 & 0 & 0 & 1
    \end{bmatrix}.
\end{align}
The corresponding eigenvalue is $\lambda_G = 0$ as the ground state is the steady-state of the system without any driving field.
Because $\rho_{L,G}$ is the identity matrix, any initial physical state $\rho_0$ projects onto the ground state: $\text{Tr}(\rho_{L,G}^\dagger\rho_0) = 1$.  
It is also the only eigenoperator with $\text{Tr}(\rho_{R,G}) = 1$.
All others are traceless $\text{Tr}(\rho_{R,\mu\neq G}) = 0$. 
In other words, the system's density matrix can be written:
\begin{align}
    \rho(t) = \rho_{R,G} + \sum_{\mu\neq G}e^{\lambda_\mu t}\text{Tr}(\rho_{L,\mu}^\dagger\rho_0)\rho_{R,\mu},
\end{align}
with the sum being traceless and always decaying to zero as $t\rightarrow\infty$.
\subsection{Symmetric and antisymmetric states}
We begin by examining the case $\gamma^* = 0$. In this regime, the symmetric and antisymmetric states, $|S\rangle$ and $|A\rangle$, each project on a single Liouvillian eigenoperator (in addition to the ground state) which we label with $\mu=S$ and $\mu=A$ respectively.
For  an initial state $\rho_0 = |S\rangle\langle S| $, the density matrix evolves as:
\begin{equation}
 \rho(t) = \rho_{R,G} + e^{\lambda_S t}\rho_{R,S}
\end{equation}
while for $\rho_0 = |A\rangle\langle A| $ one obtains:
\begin{equation}
 \rho(t) =\rho_{R,G} + e^{\lambda_A t}\rho_{R,A},
\end{equation}
with $\lambda_S = -(\Gamma_0 + \gamma_{12})$, $\lambda_A = -(\Gamma_0 - \gamma_{12})$ and:
\begin{align}
    \rho_{R,S} = \begin{bmatrix}
        -1 & 0 & 0 & 0 \\
        0 & 0.5 & 0.5 & 0 \\
        0 & 0.5 & 0.5 & 0 \\
        0 & 0 & 0 & 0
    \end{bmatrix}
    , \
    \rho_{R,A}= \begin{bmatrix}
        -1 & 0 & 0 & 0 \\
        0 & 0.5 & -0.5 & 0 \\
        0 & -0.5 & 0.5 & 0 \\
        0 & 0 & 0 & 0
    \end{bmatrix}.
\end{align}
When the dephasing rate is nonzero, $\gamma^* > 0$, both $|S\rangle\langle S|$ and $|A\rangle\langle A|$ project onto a mixture of the $\mu=A$ and $\mu = S$ eigenoperators:
\begin{align}
    \rho(t) = \rho_{R,G} + c_Ae^{\lambda_A t}\rho_{R,A} + c_Se^{\lambda_S t}\rho_{R,S},
\end{align}
with $c_\mu = \text{Tr}(\rho_{L,\mu}^\dagger\rho_0)$, $\rho_0 = |A\rangle\langle A|$ or $|S\rangle\langle S|$.
This expression explains the bi-exponential decay of the excited states populations observed in Figures~\ref{fig:g1/main}(d).
The structure of the right eigenoperators with $\gamma^* > 0$ reads:
\begin{align}
    \rho_{R,S} = \frac{1}{\gamma_{12}^*}\begin{bmatrix}
        -2\gamma_{12} & 0 & 0 & 0 \\
        0 & \gamma_{12} & \frac{1}{2}(\gamma^* + \gamma_{12}^*) & 0 \\
        0 & \frac{1}{2}(\gamma^* + \gamma_{12}^*) & \gamma_{12} & 0 \\
        0 & 0 & 0 & 0
    \end{bmatrix},\\
    \rho_{R,A} = \frac{1}{\gamma_{12}^*}\begin{bmatrix}
        -2\gamma_{12} & 0 & 0 & 0 \\
        0 & \gamma_{12} & \frac{1}{2}(\gamma^* - \gamma_{12}^*) & 0 \\
        0 & \frac{1}{2}(\gamma^* - \gamma_{12}^*) & \gamma_{12} & 0 \\
        0 & 0 & 0 & 0
    \end{bmatrix}.
\end{align}
As $\gamma^*$ increases, $\rho_{R,A}$ and $\rho_{R,S}$ evolve from representing coherent superpositions to predominantly representing populations and coherences, respectively. In the limit $\gamma^* \rightarrow \infty$ they take the simplified forms:
\begin{align}
    \rho_{R,S} \propto \begin{bmatrix}
        0 & 0 & 0 & 0 \\
        0 & 0 & 1 & 0 \\
        0 & 1 & 0 & 0 \\
        0 & 0 & 0 & 0
    \end{bmatrix}
    , \
    \rho_{R,A} \propto \begin{bmatrix}
        -1 & 0 & 0 & 0 \\
        0 & 0.5 & 0 & 0 \\
        0 & 0 & 0.5 & 0 \\
        0 & 0 & 0 & 0
    \end{bmatrix}.
\end{align}
Consequently, rather than decaying jointly through coherent coupling, the populations and coherences become dynamically decoupled under strong dephasing.
The corresponding eigenvalues are $\lambda_A = -(\Gamma_0 + \frac{1}{2}(\gamma^* - \gamma_{12}^*))$ and $\lambda_S = -(\Gamma_0 + \frac{1}{2}(\gamma^* + \gamma_{12}^*))$ which tend towards $\lambda_A \approx \Gamma_0$ and $\lambda_S \approx \Gamma_0 + \gamma^*$ for large $\gamma^*$.

The left eigenoperators are given by:
\begin{align}
    \rho_{L,S}^\dagger &= \frac{1}{\gamma_{12}^* + \gamma^*}\begin{bmatrix}
        0 & 0 & 0 & 0 \\
        0 & \gamma_{12} & \frac{(\gamma^* + \gamma_{12}^*)}{2} & 0 \\
        0 & \frac{(\gamma^* + \gamma_{12}^*)}{2} & \gamma_{12} & 0 \\
        0 & 0 & 0 & p_e
    \end{bmatrix}
    ,\\
    \text{with } p_e &= \frac{2\gamma_{12}(2\Gamma_0 + \gamma^* + \gamma_{12}^*)}{2\Gamma_0 - \gamma^* - \gamma_{12}^*}\\
    \rho_{L,A}^\dagger &= \frac{1}{\gamma_{12}^* - \gamma^*}\begin{bmatrix}
        0 & 0 & 0 & 0 \\
        0 & \gamma_{12} & \frac{(\gamma^* - \gamma_{12}^*)}{2} & 0 \\
        0 & \frac{(\gamma^* - \gamma_{12}^*)}{2} & \gamma_{12} & 0 \\
        0 & 0 & 0 & p_e
    \end{bmatrix}
    ,\\
    \text{with } p_e &= \frac{2\gamma_{12}(2\Gamma_0 + \gamma^* - \gamma_{12}^*)}{2\Gamma_0 - \gamma^* + \gamma_{12}^*}.
\end{align}
\subsection{Doubly-excited state}
The doubly-excited state $|E\rangle\langle E|$ projects onto a combination of the $\mu = G$, $A$, and $S$ liouvilian eigenstates, as well as along a fourth eigenoperator, which we denote by $\mu = E$.
As a consequence, for the initial state $\rho_0 = |E\rangle\langle E|$ the dynamics of $\rho(t)$ writes:
\begin{align}
     \rho(t) &= \rho_{R,G} + c_Se^{\lambda_S t}\rho_{R,S}\nonumber\\
    &\qquad + c_Ae^{\lambda_A t}\rho_{R,A} + c_Ee^{\lambda_E t}\rho_{R,E},
\end{align}
with
\begin{align}
    \rho_{R,E} &= \frac{1}{n}\begin{bmatrix}
        -2p -n & 0 & 0 & 0\\
        0 & p & c & 0\\
        0 & c & p & 0\\
        0 & 0 & 0 & n
    \end{bmatrix}\\
    n &= (\Gamma_0+\gamma_{12})(\Gamma_0-\gamma_{12}) - \Gamma_0\gamma^*\\
    c &= -2\Gamma_0\gamma_{12}\\
    p &= \Gamma_0\gamma^* - \Gamma_0^2 - \gamma_{12}^2.
\end{align}
Similarly to $\rho_{R,A}$, the coherences in $\rho_{R,E}$ vanish as $\gamma^*$ increases and the decay of the doubly excited state ceases to create coherences.
The corresponding left eigenoperator is:
\begin{align}
    \rho_{L,E}^\dagger = \begin{bmatrix}
        0 & 0 & 0 & 0 \\
        0 & 0 & 0 & 0 \\
        0 & 0 & 0 & 0 \\
        0 & 0 & 0 & 1
    \end{bmatrix}.
\end{align}
The remaining twelve eigenvalues primarily describe the evolution of the off-diagonal elements of the density matrix.
These modes are particularly relevant for the computation of two-time correlation functions $G^{(1)}(t, t+\tau)$ and their Fourier transforms.
\section{Expectation value and correlation functions using the spectral decomposition of the Liouvillian}
In this appendix, we show how the spectral decomposition of the Liouvillian provided in eq. (\ref{eq:rhoLiouv}) permits to compute the expectation value of different observables.
We can compute the time resolved expectation value of $\mathcal{O}$ using the expression:
\begin{eqnarray}
    \langle\mathcal{O}(t)\rangle &&= \text{Tr}\left(\mathcal{O}\rho(t)\right)\nonumber \\ 
    &&= \sum_\mu e^{\lambda_\mu t}\text{Tr}\left(\rho^\dagger_{L\mu}\rho_0\right)\text{Tr}\left(\mathcal{O}\rho_{R,\mu}\right).
\end{eqnarray}
This expression can be interpreted as follows: the expectation value $ \langle\mathcal{O}(t)\rangle $ is given by the sum over all eigenoperators of $\mathcal{L}$ , each term being the projection of the initial state onto the corresponding left eigenoperator, multiplied by the expectation value of $\mathcal{O}$ evaluated on the associated right eigenoperator. 

The correlation functions between any two operators $\mathcal{A}$ and $\mathcal{B}$ can be computed from the quantum regression theorem:
\begin{align}
\langle\mathcal{A}(t)\mathcal{B}(t+\tau)\rangle
&= \text{Tr}\left(\mathcal{B}e^{\mathcal{L\tau}}[\rho(t)\mathcal{A}]\right) \\
&= \sum_{\mu, \mu'}e^{\lambda_\mu t}e^{\lambda_{\mu'} \tau}
\text{Tr}\left(\rho^\dagger_{L,\mu}\rho_0\right) \label{eq:quantum_reg_2_op}\\
&\quad \times
\text{Tr}\left(\rho^\dagger_{L,\mu'}\rho_{R,\mu}\mathcal{A}\right)
\text{Tr}\left(\mathcal{B}\rho_{R,\mu'}\right).
\end{align}

This expression can be interpreted as follows: the correlation function  $ \langle\mathcal{A}(t)  \mathcal{B}(t+\tau)\rangle $ is expressed as a sum over pairs ($\mu,\mu'$) of eigenoperators of $\mathcal{L}$. Each term  of the sum corresponds to the projection of the initial state  onto the first left eigenoperator $\mu$. The operator $\mathcal{A}$ then couples ($\mu,\mu'$), and the contribution is finally weighted by the expectation value of $\mathcal{B}$ evaluated on the right eigenoperator associated with $\mu'$.

The same reasoning applies to correlation between any 3 operators $\mathcal{A}$, $\mathcal{B}$ and $\mathcal{C}$:
\begin{eqnarray}
   && \langle\mathcal{A}(t)\mathcal{B}(t+\tau)\mathcal{C}(t)\rangle = \sum_{\mu, \mu'}e^{\lambda_\mu t}e^{\lambda_{\mu'} \tau}\label{eq:quantum_reg_3_op}\\
&&\times\text{Tr}\left(\rho^\dagger_{L,\mu}\rho_0\right)\nonumber\text{Tr}\left(\rho^\dagger_{L,\mu'}\mathcal{C}\rho_{R,\mu}\mathcal{A}\right)\text{Tr}\left(\mathcal{B}\rho_{R,\mu'}\right),\nonumber
\end{eqnarray}
with a similar interpretation: the correlation function  $  \langle\mathcal{A}(t)\mathcal{B}(t+\tau)\mathcal{C}(t)\rangle$ is expressed as a sum over pairs ($\mu,\mu'$) of eigenoperators of $\mathcal{L}$. Each term  of the sum corresponds to the projection of the initial state  onto the first left eigenoperator $\mu$. The operators $\mathcal{A}$ and $\mathcal{C}$ then couples ($\mu,\mu'$), and the contribution is finally weighted by the expectation value of $\mathcal{B}$ evaluated on the right eigenoperator associated with $\mu'$.
Applying this to the expression of $g^{(2)}(t,t+\tau)$ provided in eq. (\ref{eq:g2_t_tau}) yields:
\begin{eqnarray}
        \label{eq:g2_t_tau_eigendecomposition}
        g^{(2)}&(t, t+\tau) = \Bigl\{ \sum_{\mu,\mu'} e^{\lambda_\mu t}e^{\lambda_{\mu'}\tau}
        \text{Tr}\left(\rho^\dagger_{L,\mu}\rho_0\right)\\
        &\times\text{Tr}\left(\rho^\dagger_{L,\mu'}D\rho_{R,\mu}D^\dagger\right)
        \text{Tr}\left(D^\dagger D\rho_{R,\mu'}\right)\Bigr\}\\      
        &/\Bigl(\langle\mathcal{D^\dagger D}(t)\rangle\langle\mathcal{D^\dagger D}(t+\tau)\rangle          
            \Bigr)
\end{eqnarray}
with
\begin{equation}
\langle\mathcal{D^\dagger D}(t+\tau)\rangle =
            \sum_\mu e^{\lambda_\mu (t+\tau)}
            \text{Tr}\left(\rho^\dagger_{L,\mu}\rho_0\right)
            \text{Tr}\left(D^\dagger D\rho_{R,\mu}\right)
\end{equation}
Out of the 256 possible combinations in the double sum, only a few yield non-zero contributions.  Considering that $\rho_0=\vert E\rangle \langle E \vert$, we first evaluate $\text{Tr}\left(\rho^\dagger_{L,\mu}\rho_0\right)$ which selects only 4 Liouvillian eigenoperators corresponding to $\mu$ =($E, S, A$ and $G$):
\begin{align}
    \text{Tr}\left(\rho^\dagger_{L,E}\rho_0\right) &= 1 \\
    \text{Tr}\left(\rho^\dagger_{L,S}\rho_0\right) &= 2\gamma_{12} \frac{2\Gamma_0 + \gamma^* + \gamma^*_{12}}{(\gamma^* + \gamma^*_{12})(2\Gamma_0 - \gamma^* - \gamma^*_{12})}\\
    \text{Tr}\left(\rho^\dagger_{L,A}\rho_0\right) &= 2\gamma_{12} \frac{2\Gamma_0 + \gamma^* - \gamma^*_{12}}{(-\gamma^* + \gamma^*_{12})(2\Gamma_0 - \gamma^* + \gamma^*_{12})}\\
    \text{Tr}\left(\rho^\dagger_{L,G}\rho_0\right) &= 1,
\end{align}
where we have defined $\gamma_{12}^* = \sqrt{{\gamma^*}^2 + 4\gamma_{12}^2}$. 
Then, the term $\text{Tr}\left(D^\dagger D\rho_{R,\mu'}\right)$ restricts the sum to three values of $\mu'$, namely $E$, $S$, and $A$. Using the collective dipole operator $D = (e^{i\phi/2}\sigma_1 + e^{-i\phi/2}\sigma_2)/\sqrt{2}$
we obtain:
\begin{align}
    \text{Tr}\left(D^\dagger D\rho_{R,E}\right) &= -2\gamma_{12}\frac{\Gamma_0\cos\phi + \gamma_{12}}{\Gamma_0^2 - \gamma_{12}^2 - \Gamma_0\gamma^*}\\
    \text{Tr}\left(D^\dagger D\rho_{R,S}\right) &= \frac{2\gamma_{12} + (\gamma^* + \gamma_{12}^*)\cos\phi}{2\gamma_{12}^*}\\ 
    \text{Tr}\left(D^\dagger D\rho_{R,A}\right) &= \frac{2\gamma_{12} + (\gamma^* - \gamma_{12}^*)\cos\phi}{2\gamma_{12}^*}
\end{align}
Consequently, only 12  $(\mu, \mu')$ combinations need to be evaluated in the term: $\text{Tr}\left(\rho^\dagger_{L,\mu'}D\rho_{R,\mu}D^\dagger\right)$. Among these, only the pairs $(E,S)$ and $(E,A)$ contribute to nonzero values:
\begin{align}
    \text{Tr}\left(\rho^\dagger_{L,S}D\rho_{R,E}D^\dagger\right) &= \frac{2\gamma_{12} + (\gamma^* +  \gamma^*_{12})\cos\phi}{2(\gamma^*_{12} + \gamma^*)}\\
    \text{Tr}\left(\rho^\dagger_{L,A}D\rho_{R,E}D^\dagger\right) &= \frac{2\gamma_{12} + (\gamma^* -  \gamma^*_{12})\cos\phi}{2(\gamma^*_{12} - \gamma^*)}.
\end{align}
This allows a drastic simplification of the numerator in Eq.~(\ref{eq:g2_t_tau_eigendecomposition}), which reduces to:
\begin{widetext}
    \begin{align}
        G^{(2)}(t,t+\tau) &= e^{\lambda_E t} \text{Tr}\left(\rho^\dagger_{L,E}\rho_0\right) \left(
        e^{\lambda_{S} \tau} 
        \text{Tr}\left(\rho^\dagger_{L,S}D\rho_{R,E}D^\dagger\right)
        \text{Tr}\left(D^\dagger D\rho_{R,S}\right) +
        e^{\lambda_{A} \tau} 
        \text{Tr}\left(\rho^\dagger_{L,A}D\rho_{R,E}D^\dagger\right)
        \text{Tr}\left(D^\dagger D\rho_{R,A}\right)\right) \nonumber\\
        &= e^{-2\Gamma_0 t}e^{-(\Gamma_0 + \frac{1}{2}\gamma^*)\tau}\left(
        e^{-\gamma^*_{12}\tau/2}\frac{(2\gamma_{12} + (\gamma^* + \gamma^*_{12})\cos\phi)^2}{4\gamma^*_{12}(\gamma^*_{12} + \gamma^*)}
        +
        e^{+\gamma^*_{12}\tau/2}\frac{(2\gamma_{12} + (\gamma^* - \gamma^*_{12})\cos\phi)^2}{4\gamma^*_{12}(\gamma^*_{12} - \gamma^*)}
        \right).
    \end{align}
\end{widetext}
\section{\label{app:The baths}The different baths.}
In this appendix, we provide further details on the properties of the different baths introduced in our model, namely the photonic bath and the dephasing baths. We first discuss the temperature of the photonic bath and justify why it can generally be treated as being at zero temperature for optical photons. We then consider the dephasing baths and show that, for experimentally relevant parameters, they typically operate in the high-temperature regime with respect to the energy scale set by the dipole-dipole interaction. We next discuss the microscopic form of the system-bath coupling and motivate the use of a linear interaction for the effective description considered in this work. For completeness, we then investigate the global master equation in the low-temperature regime, where it differs significantly from the local description, and illustrate its consequences both in the absence of a coherent drive and in the presence of a weak coherent drive.
\subsection*{The photonic bath is effectively at $0$K}
We now discuss the temperature of the photonic bath, which can be associated with the thermal electromagnetic field, or black-body radiation, at temperature $T$.
The relevant quantity is the mean thermal photon occupation number at frequency $\omega_0$ associated to the atomic transition,
\begin{equation}
n_{\text{th}}(\omega_0) = \frac{1}{e^{\hbar \omega_0 / k_B T} - 1}.
\end{equation}
For optical transitions, $\hbar \omega_0$ typically lies in the eV range.
Even at room temperature,
\begin{equation}
k_B T (300\,\text{K}) \approx 2.6 \times 10^{-2} \, \text{eV},
\end{equation}
so that $\hbar \omega_0 \gg k_B T$ for optical photons.
As a consequence,
\begin{equation}
n_{\text{th}}(\omega_{0}) \approx 0,
\end{equation}
and the photonic bath can be considered effectively at zero temperature.
In this regime, the bath induces spontaneous emission but no thermally activated absorption processes.
Thermal photon effects would become relevant only if $\hbar \omega_0$ were comparable to $k_B T$, which requires much lower frequencies.
At room temperature, this condition corresponds to the terahertz (THz) range, where $\hbar \omega_0 \sim k_B T$.
In that regime, the thermal photon occupation number is no longer negligible and the photonic bath can no longer be treated as being at $0$K.
Since the emitters considered in the present work operate in the optical domain, the zero-temperature approximation for the photonic bath is fully justified. 
\subsection*{The dephasing bath is generally in the high-temperature regime}
\begin{figure}[h]
\centering
\includegraphics[width=0.45\textwidth]{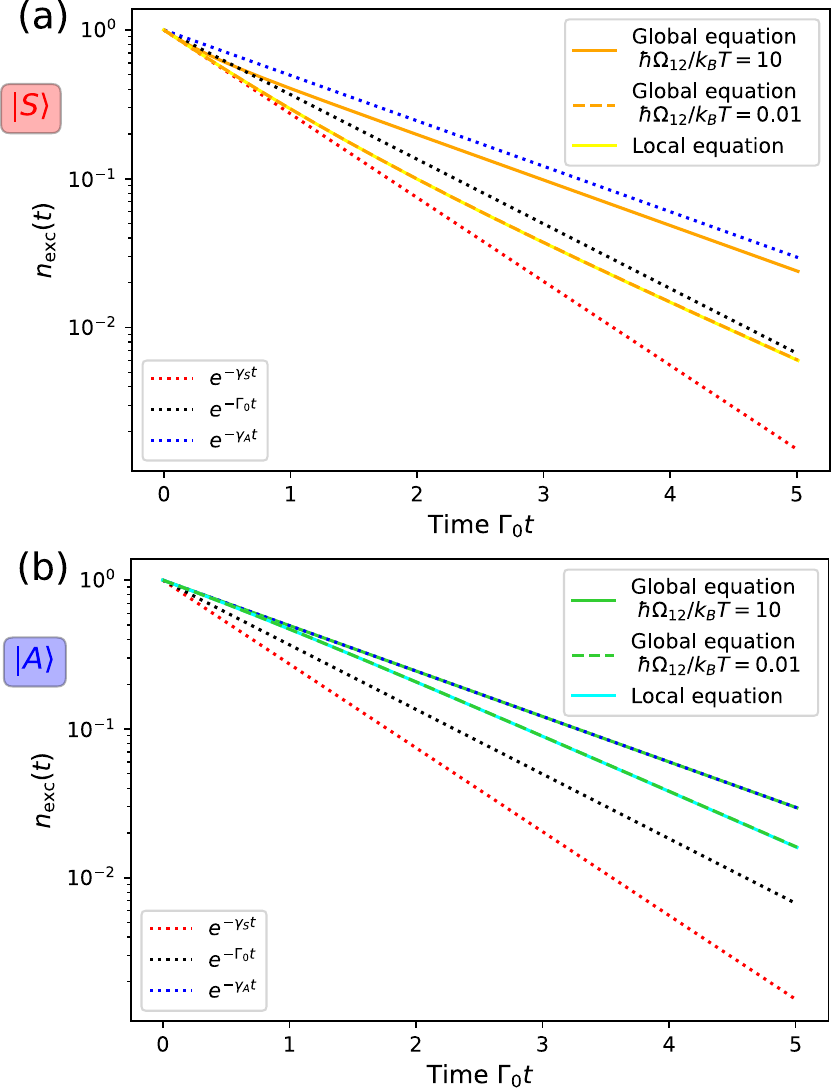}
\caption{\label{fig:app_global_vs_local_relaxation}Relaxation dynamics of the system predicted by the local equation and the global equation for $\hbar \Omega_{12}/k_B T = 0.01$ (high temperature regime) and $\hbar \Omega_{12}/k_B T = 10$ (low temperature regime). We plot the total number of excitations $n_\text{exc}(t)$ starting from the symmetric state $|S\rangle$ (a) and from the anti-symmetric state $|A\rangle$ (b) for a fixed dephasing rate $\gamma^* = 0.5 \Gamma_0$. The emitters are in the same configuration as in Figs. \ref{fig:g1/main} (a) and (d) of the main text such that the dipole-dipole coupling is relatively strong with $\Omega_{12} = 20 \Gamma_0$. To better visualize the effect of dephasing, we compare these curves to the superradiant, subradiant, and independent-emitter decay in the absence of dephasing (dotted colored lines).} 
\end{figure}

As discussed in Sec.~\ref{sec_deph_coupling}, the dipole-dipole interaction splits the symmetric and antisymmetric collective states $\lvert S\rangle$ and $\lvert A\rangle$ by an energy $2\hbar\Omega_{12}$. The relevant temperature scale for the dephasing bath is therefore set by the ratio $\hbar\lvert \Omega_{12}\rvert/k_B T$. For experimentally relevant parameters, this ratio is typically much smaller than unity. For instance,
\begin{align}
k_B T (1\,\text{K}) &\approx 8.6 \times 10^{-5} \, \text{eV}, \\
k_B T (300\,\text{K}) &\approx 2.6 \times 10^{-2} \, \text{eV}.
\end{align}
In contrast, $\hbar \Omega_{12}$ in typical emitter systems lies in the $\mu\text{eV}$ range. For example, taking $\Gamma_0/2\pi = 60\,\text{MHz}$ and $\Omega_{12}=20\Gamma_0$ gives $\hbar\Omega_{12}\simeq 5\,\mu\text{eV}$, and therefore
\begin{equation}
\frac{\hbar\Omega_{12}}{k_B T}\ll 1
\end{equation}
even at $T=1\,\text{K}$. Experimental values reported for molecular emitters can be even smaller, with $\hbar\lvert \Omega_{12}\rvert$ of the order of $1\,\mu\text{eV}$~\cite{Trebbia2022,Lange2024}.
In this regime, the dephasing bath effectively behaves as an infinite-temperature bath with respect to the energy splitting between $\lvert S\rangle$ and $\lvert A\rangle$. Consequently, thermally induced upward and downward transitions between these states occur with nearly equal rates, $\Gamma_{+-}(T)\simeq\Gamma_{-+}(T)\simeq\gamma^*(T)$. The bath can therefore both absorb and supply energy on the scale $\hbar\lvert \Omega_{12} \rvert$, allowing population transfer in both directions while remaining fully consistent with thermal equilibrium.
This high-temperature regime also provides the physical justification for the local master equation used throughout the main text. As discussed in Sec.~\ref{sec_deph_coupling}, the global description accounts explicitly for the energy splitting between the collective states, whereas the local description neglects it in the derivation of the dephasing contribution. In the limit $\hbar\lvert \Omega_{12} \rvert/k_B T\rightarrow 0$, the two descriptions become equivalent, since $\Gamma_{+-}(T)\simeq\Gamma_{-+}(T)\simeq\gamma^*(T)$. This convergence is illustrated in Fig.~\ref{fig:app_global_vs_local_relaxation}, which compares the relaxation dynamics predicted by the local and global master equations for $\hbar\Omega_{12}/k_B T=0.01$~\cite{shishkov2019relaxation}. The figure also illustrates that the local approximation remains highly accurate for experimentally relevant values of this ratio. 
\subsection*{The linear coupling between the system and the dephasing bath}
The dephasing bath is coupled to the system via
\begin{equation}
H_{\text{mat-B}i} = \hbar \sum_{k} \gamma_{k,i} \, \sigma_i^{ee} \left(b_{k,i}^{\dagger} + b_{k,i}\right).
\end{equation}
This corresponds to a linear system--bath coupling, which represents the minimal microscopic mechanism leading to pure dephasing within a standard Born--Markov treatment.
More generally, the interaction Hamiltonian could also contain quadratic terms of the form
\begin{equation}
H_{\text{quad}} = \hbar \sum_{k,k'} \eta_{k k',i} \, \sigma_i^{ee} 
\left(b_{k,i}^{\dagger} + b_{k,i}\right)
\left(b_{k',i}^{\dagger} + b_{k',i}\right),
\end{equation}
which describe two-phonon processes \cite{clear2020phonon}. 
Such contributions can become relevant in a more microscopic description of emitter-phonon interactions and may lead to additional temperature-dependent corrections to the effective dephasing rate. However, the aim of this work is not to model the microscopic origin of pure dephasing or its detailed temperature dependence, but rather to investigate how an effective dephasing channel competes with collective effects. For this purpose, the linear coupling provides a minimal description that captures the relevant pure-dephasing mechanism while allowing the temperature dependence to be incorporated into the effective rate $\gamma^*(T)$.
\subsection*{Global equation in the low temperature regime}
From Eq.~\eqref{eq:global_fct_local}, the difference between the global and local descriptions becomes increasingly pronounced as the ratio $\hbar \lvert \Omega_{12} \rvert/k_B T$ increases. 
In the low-temperature regime $\hbar \lvert \Omega_{12} \rvert/k_B T \gg 1$, thermally activated transitions towards the higher-energy collective state are suppressed, so that population transfer between $\lvert A\rangle$ and $\lvert S\rangle$ occurs predominantly towards the lower-energy state.
For $\Omega_{12}>0$, the antisymmetric state $\lvert A\rangle$ is the lower-energy state, and the transition rates scale as $\Gamma_{-+}\simeq 0$ and $\Gamma_{+-}(T)~\sim~\gamma^*(T)\hbar \vert\Omega_{12}\vert/k_B T$. Conversely, for $\Omega_{12}<0$, the symmetric state $\lvert S\rangle$ is the lower-energy state, and $\Gamma_{+-}\simeq 0$ and $\Gamma_{-+}(T)~\sim~\gamma^*(T)\hbar \vert \Omega_{12}\vert/k_B T$. In this regime, the dephasing-induced population transfer between the collective states can therefore dominate over the residual diagonal dephasing dynamics. In the absence of driving and detuning ($\Omega_R=\Delta=0$), the system consequently relaxes at long times towards the subradiant state for $\Omega_{12}>0$ and towards the superradiant state for $\Omega_{12}<0$.
We illustrate this behavior for $\Omega_{12}/\Gamma_0=20$ in Fig.~\ref{fig:app_global_vs_local_relaxation}, where we compare the excited-state population $n_\text{exc}(t)$ obtained from the local and global master equations for $\hbar\Omega_{12}/k_B T=10$. Starting from the symmetric state $\lvert S\rangle$, the dynamics initially exhibits superradiant decay at a rate $\gamma_S=\Gamma_0+\gamma_{12}$, followed by a transfer towards the antisymmetric state and subsequent subradiant decay at a rate $\gamma_A=\Gamma_0-\gamma_{12}$. In contrast, when the system is initially prepared in the antisymmetric state $\lvert A\rangle$, the decay remains subradiant throughout the evolution. This behavior differs qualitatively from the local description, in which pure dephasing induces transitions between the collective modes and thereby tends to wash out their distinct radiative character. In the global description, by contrast, low-temperature dephasing preferentially transfers population towards the lower-energy collective mode, thereby preserving and even favoring either subradiant or superradiant emission depending on the sign of $\Omega_{12}$. This behavior suggests that temperature-dependent dephasing could potentially be exploited to control the population and emission dynamics of collective modes.
\subsection*{Global equation with a coherent drive}
\begin{figure}[h]
\centering
\includegraphics[width=0.45\textwidth]{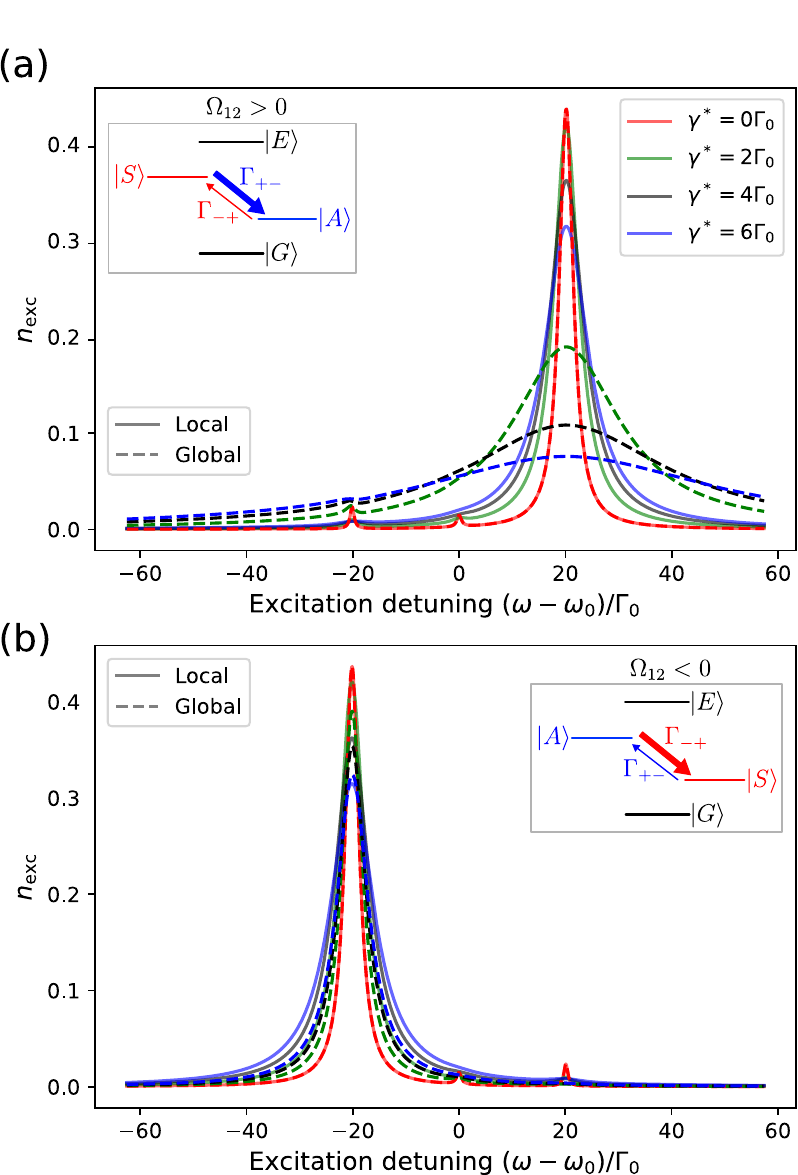}
\caption{\label{fig:exc_spectrum_global_vs_local} Excitation spectrum for different dephasing rates calculated with the stationary solutions of the local equation (full colored lines) and the global equation (dashed lines) for $\hbar \lvert \Omega_{12} \rvert/k_B T = 10$. In both simulations, the Rabi frequency is $\Omega_R = 1.5 \Gamma_0$ and the frequency detuning is $\Delta = 5\Gamma_0$. (a) Results for emitters in the 'H' configuration where the dipole moments are orthogonal to the emitter axis. The inter-emitter spacing is $r \simeq 0.035 \lambda$ and the Debye-Waller/Franck-Condon factor is $\alpha = 0.3$, such that the dipole-dipole coupling is $\Omega_{12} = 20 \Gamma_0$. (b) Results in the 'J' configuration where the dipole moments are parallel to the emitter axis. The inter-emitter spacing is $r \simeq 0.0455 \lambda$ and the Debye-Waller/Franck-Condon factor is $\alpha = 0.3$, such that the dipole-dipole coupling is $\Omega_{12} = -20 \Gamma_0$.}
\end{figure}

Let us now consider the effect of a coherent drive within the global master-equation framework. Here, the drive is introduced phenomenologically through the system Hamiltonian, while the dissipative terms are kept unchanged from the undriven global master equation. A fully consistent derivation would require reapplying the Davies procedure to the driven system, whose eigenstates and transition frequencies are modified by the coherent drive. Such a treatment is beyond the scope of the present work. We instead focus on the weak-drive regime $\Omega_R\ll\lvert\Omega_{12}\rvert$, where the collective basis ${|G\rangle,|E\rangle,|A\rangle,|S\rangle}$ remains a useful basis for describing the dephasing dynamics.
For weak drive amplitudes, the population and coherences involving the doubly excited state $|E\rangle$ can be neglected. We can therefore restrict the dynamics to the ground state and the single-excitation manifold. In this approximation, the global dynamics are described by:
\begin{align}
	\frac{d\rho_{A,A}}{dt} &= -\left(\gamma_A + \Gamma_{-+}/2 \right) \rho_{A,A} + \left(\Gamma_{+-}/2\right) \rho_{S,S} \,\,, \label{global_eqs_1exc_1}\\
	\frac{d\rho_{S,S}}{dt}&= -\left(\gamma_S  + \Gamma_{+-}/2\right)\rho_{S,S}+ \left(\Gamma_{-+}/2\right)\rho_{A,A} \notag  \\
    &+ \sqrt{2}\,\Omega_R \text{Im}(\rho_{G,S}),\label{global_eqs_1exc_2}\\
	\frac{d\rho_{G,S}}{dt} &= i\left(\Omega_R/\sqrt{2}\right) (1 - \rho_{A,A} - 2\rho_{S,S}) \notag \\
    &+ \left[i(\delta + \Omega_{12}) - (\gamma^*/4 + \Gamma_{+-}/4 +\gamma_S/2)\right]\rho_{G,S}, \label{global_eqs_1exc_3}
\end{align}
where $\delta = \omega - \omega_0$ is the laser detuning. In the high temperature regime $\hbar \lvert \Omega_{12}\rvert/k_B T \ll 1$, $\Gamma_{+-} \simeq \Gamma_{-+} \simeq \gamma^*$ and these equations coincide with their local counterpart. 
More interestingly, in the low-temperature regime $\hbar|\Omega_{12}|/k_B T\gg1$, the effect of dephasing depends on whether the superradiant state $|S\rangle$ is the higher or lower-energy collective state. This behavior is illustrated in Fig.~\ref{fig:exc_spectrum_global_vs_local}, where we compare the excitation spectra obtained from the local and global equations for different dephasing rates. The excitation spectrum is dominated by the superradiant resonance, making it particularly sensitive to whether dephasing can induce population transfer out of the superradiant mode.
For $\Omega_{12}>0$ [Fig.~\ref{fig:exc_spectrum_global_vs_local}(a)], the superradiant state $|S\rangle$ is the higher-energy state. At low temperature, dephasing therefore induces a strong population transfer from $|S\rangle$ to the lower-energy subradiant state $|A\rangle$, with $\Gamma_{+-}\gg\gamma^*$. This additional decay channel broadens and depletes the superradiant resonance, while slightly enhancing the subradiant contribution. Conversely, for $\Omega_{12}<0$ [Fig.~\ref{fig:exc_spectrum_global_vs_local}(b)], the superradiant state $|S\rangle$ is the lower-energy state. The reverse transition $|S\rangle\rightarrow|A\rangle$ is then thermally suppressed, with $\Gamma_{+-}\simeq0$. As a result, the superradiant state is largely protected from dephasing-induced population transfer, and the corresponding resonance is slightly more robust against dephasing than in the local description.
The different behaviors can also be understood directly from Eq.~\eqref{global_eqs_1exc_3}. The damping rate of the coherence $\rho_{G,S}$, which describes the coupling between the ground state and the superradiant state, is given by $\gamma_{gs}=\gamma^*/4+\Gamma_{+-}/4+\gamma_S/2$. When $\Omega_{12}>0$, the superradiant state $|S\rangle$ is the higher-energy state and $\Gamma_{+-}$ is large at low temperature. The resulting additional contribution $\Gamma_{+-}/4$ therefore enhances the damping of $\rho_{G,S}$, leading to a stronger broadening and depletion of the superradiant resonance. Conversely, when $\Omega_{12}<0$, $|S\rangle$ is the lower-energy state and $\Gamma_{+-}\simeq0$, so that this additional damping channel is absent and the superradiant resonance is comparatively protected from dephasing.
These results provide a simple physical picture of the temperature-dependent behavior of the excitation spectrum: at low temperature, dephasing strongly affects the superradiant resonance when the latter is the higher-energy collective state, whereas it has a much weaker effect when the superradiant state is the lower-energy state.

\clearpage
\bibliography{biblio}

\end{document}